\documentclass[12pt]{article}
\usepackage{graphicx,psfrag,epsf}
\usepackage{enumerate}
\usepackage{natbib}
\usepackage{moreverb,url,multirow}
\usepackage{xcolor}
\usepackage{cprotect}
\usepackage{float}
\usepackage[caption=false]{subfig}

\usepackage{amsmath,amsfonts,amssymb,amsthm}
\usepackage[algosection,linesnumbered,lined,boxed,commentsnumbered]{algorithm2e}

\usepackage{xr}
\usepackage{xcite}
\externalcitedocument{supplement}

\def\mcL{\mathcal{L}}
\def\mcT{\mathcal{T}}

\newcommand{\tr}{\operatorname{tr}} % Trace

\def\bA{\mathbf{A}}
\def\bG{\mathbf{G}}

\def\bY{\mathbf{Y}}

\def\bG{\mathbf{G}}

\def\bY{\mathbf{Y}}

\def\bpsi{\boldsymbol{\psi}}

\def\bmu{\boldsymbol{\mu}}

\theoremstyle{plain}
\newtheorem{thm}{Theorem}

\newtheorem{prop}[thm]{Proposition}

\newcommand{\blind}{0}

\begin{document}

\def\spacingset#1{\renewcommand{\baselinestretch}%
{#1}\small\normalsize} \spacingset{1}

%%%%%%%%%%%%%%%%%%%%%%%%%%%%%%%%%%%%%%%%%%%%%%%%%%%%%%%%%%%%%%%%%%%%%%%%%%%%%%

\if0\blind
{
  \title{\bf Power and sample size calculation of two-sample projection-based testing for sparsely observed functional data}
  \author{Salil Koner\thanks{
    Corresponding author. Salil Koner is postdoctoral researcher at the department of Biostatistics and Bioinformatics of Duke University. Email: salil.koner@duke.edu.} 
    \;and 
    Sheng Luo\thanks{
    Sheng Luo is a professor at the Department of Biostatistics and Bioinformatics of Duke University. Email: sheng.luo@duke.edu.
    The authors gratefully acknowledge the support from National Institute of Aging (R01AG064803, P30AG072958, and P30AG028716).} \\ 
    Department of Biostatistics and Bioinformatics, Duke University}
  \maketitle
} \fi

\if1\blind
{
  \bigskip
  \bigskip
  \bigskip
  \begin{center}
    {\LARGE\bf Power and sample size calculation for two-sample projection-based testing of sparsely observed functional data}
\end{center}
  \medskip
} \fi

\begin{abstract}
Projection-based testing for mean trajectory differences in two groups of irregularly and sparsely observed functional data has garnered significant attention in the literature because it accommodates a wide spectrum of group differences and (non-stationary) covariance structures. This article presents the derivation of the theoretical power function and the introduction of a comprehensive power and sample size (PASS) calculation toolkit tailored to the projection-based testing method developed by \cite{wang2021two}. Our approach accommodates a wide spectrum of group difference scenarios and a broad class of covariance structures governing the underlying processes. Through extensive numerical simulation, we demonstrate the robustness of this testing method by showcasing that its statistical power remains nearly unaffected even when a certain percentage of observations are missing, rendering it ``missing-immune''. Furthermore, we illustrate the practical utility of this test through analysis of two randomized controlled trials of Parkinson's disease. To facilitate implementation, we provide a user-friendly R package \verb|fPASS|, complete with a detailed vignette to guide users through its practical application. We anticipate that this article will significantly enhance the usability of this potent statistical tool across a range of biostatistical applications, with a particular focus on its relevance in the design of clinical trials.
\end{abstract}

\noindent%

\spacingset{1.45}

\section{Introduction} 
The comparison of two groups based on their longitudinal data plays a pivotal role in various practical applications, especially in the context of clinical trials. To assess the efficacy of study drugs or interventions, statistical methods for two-sample testing are commonly employed, with mixed models and generalized estimating equations (GEE) being popular choices in modern clinical trials \citep{diggle2002analysis}. These methodologies offer numerous advantages, including extensive research support and user-friendly statistical software for implementation. They also facilitate power and sample size calculations, making them valuable tools in clinical trial design \citep{brown2015applied}. However, these methods have certain limitations, as they rely on restrictive assumptions on the evolution of mean trajectory and require the specification of stationary covariance structures as in GEE, which can become computationally intensive with multiple (more than three) random effects. Additionally, many tests demand uniform observation times and regular intervals between subsequent visits for each subject, impacting their applicability in many situations.

% Recent statistical advancements have introduced robust testing procedures for longitudinal data analysis, particularly in the context of sparse and irregular functional data.

Alternatively, one can consider treating longitudinal outcome measurments as sparsely observed functional data and employ flexible data-adaptive nonparametric approaches tailored to functional data analysis. These procedures allow for nonparametric evolution of mean functions over time, eliminate the need for rigid covariance structure assumptions \citep{guo2002functional, yao2005functional}. Functional data observed under a sparse and irregular design pose distinct challenges due to limited subject-specific observations. In such scenarios, these test procedures leverage the smoothness of temporal mean and covariance functions, aggregating information across subjects to infer underlying mean trajectories. Tests for sparse functional data can be categorized into two main groups. The first group includes likelihood ratio tests \citep{crainiceanu2004likelihood} and their various extensions to complex covariance structures \citep{wang2012testing, staicu2014likelihood}. The second group involves projecting data onto data-driven leading directions and using projection scores to test differences between the two groups. \cite{pomann2016two} introduced an Anderson-Darling statistic for testing equality in distributions using the projection scores, while \cite{wang2021two} proposed a Hotelling's $T^2$ type statistic to test mean differences based on these scores. The advantage of the Hotelling $T^2$ test lies in its ability to assess group differences using a single statistic, whereas the other  requires multiple comparisons. Furthermore, the Hotelling $T^2$ type test can be extended to accommodate multiple groups. Despite their flexibility, these tests have mainly been confined to statistical methodology researchers due to the lack of established finite-sample power function formulas and sample size calculation software. Consequently, they have not been widely adopted in clinical trial design practices or as primary statistical approaches in the statistical analysis plans (SAP) for clinical trials.

In this article, we derive a sample size and power calculation (PASS) formula for \cite{wang2021two}'s projection-based Hotelling $T^2$ type test designed for sparse and irregularly observed functional data. We begin by deriving the test's theoretical power function and provide a formula to determine the minimum sample size required to achieve a specified level of statistical power by inverting the power function. The test statistic's non-null distribution approximately follows a non-central chi-squared distribution, with the non-centrality parameter comprising the projection of the true mean difference onto the principal directions. Our PASS formula is highly flexible, accommodating various smooth group difference structures without requiring predefined parametric forms like linear or quadratic functions. Users can specify non-stationary smooth covariance structures for the response, alongside standard parametric structural covariances used in longitudinal data analysis. Additionally, our formula accommodates varying observation numbers and irregular schedules for each subject.

The performance of the test's power function relies on accurate estimation of the covariance structure and data-driven directions. Numerical studies demonstrate that, with a reasonably large sample size, the test's statistical power remains robust even with an increasing percentage of missing observations per subject within a certain range. This property makes the test ``missing-immune,'' particularly valuable in real-world scenarios with common missing data issues. To illustrate practical utility, we apply the test to two clinical trial examples involving Parkinson's Disease (PD) patients, showing its potential to reduce the required number of subjects to achieve specified statistical power levels as outlined in the respective statistical analysis plan (SAP) documents. To enhance the practical applicability of this testing mechanism in clinical trial simulation, we have developed a user-friendly R package \verb|fPASS| \citep{fPASS2023}. This package facilitates power analysis for various covariance structures in the analysis of longitudinal responses, along with arbitrary effect size functions, making the projection-based test more accessible to practitioners in clinical trial design and analysis.

The remainder of this article is organized as follows: Section~\ref{sec: testingframework} provides a concise description of the testing procedure and presents the non-null distribution of the test statistic. Section~\ref{sec: PASS} details the PASS formula. Validation of the formula is demonstrated through various numerical studies in Section~\ref{sec: edfd_testing_simstudy}, while real-data applications are presented in Section~\ref{sec: realdata}. Proofs of theorems are included in Section~\ref{sec: proof_of_theorems} of Supplementary material. R codes for the simulation studies are available at \url{https://github.com/SalilKoner/ufPASS}.

\label{sec: edfd_testing_intro}

\section{Projection-based testing framework}  \label{sec: testingframework}

%\subsection{Projection-based testing procedure of \cite{wang2021two}}\label{sec: edfd_testing_Statframework}

We first give a brief overview of the projection-based testing framework developed in \cite{wang2021two}. Let $Y_{ij,g}$ denotes the $j$th observation for the $i$th subject in the $g$th group, for $g=1,2$, $i=1, \ldots, n_g$, and $j=1, \dots, N_{i,g}$, with $n =n_1 + n_2$ being the total number of subjects combining the two groups. Let $T_{ij, g}$ is the associated time point when $Y_{ij,g}$ is observed. We posit the following model for the response, 
%We assume that the observed data is a noisy realization of an underlying stochastic process $X_g(t) \in \mcL^2(\mcT)$, the Hilbert space of all square integrable random functions in $\mcT$ a compact set in $\Real$, as 
\begin{equation} \label{eqn: dgm}
    Y_{ij,g} = X_{i,g}(T_{ij, g}) +  \epsilon_{ij,g},
\end{equation} 
where $X_{i,g}(\cdot)$ is the latent response trajectory and $\epsilon_{ij, g}$ is the measurement error. For each group $g$, we assume that the latent process $\{X_{i,g}(t)\}_{i=1}^{n_g}$ are independent and identically distributed (i.i.d) copies of an underlying stochastic process $X_g(t) \in \mathcal{L}^2(\mcT)$, the space of all square integrable functions in the compact domain $\mcT$, equipped with the inner product $\langle f_1, f_2 \rangle:= \int_{\mcT} f_1(t)f_2(t)\,dt$, for  $f_1(t), f_2(t) \in \mathcal{L}^2(\mcT)$. We assume that $X_g(t)$ have a group specific mean function $\mu_g(t)$ and a (common) covariance operator $(\Xi\;f)(t) = \int \Sigma(t,s)f(s)ds$, induced by the covariance kernel $\Sigma(t,s) = \text{Cov}(X_g(t), X_g(s))$. The latent trajectories are assumed to be independent across the groups as well. Lastly, measurement errors $\{\epsilon_{ij, g}\}$ are assumed to be i.i.d across all indices with mean zero and variance $\tau^2$. Encompassing the random sampling design \citep{yao2005functional}, we assume that the number of observations for each subject, $N_{i,g}$ is small and finite and the observation points $\{T_{ij,g}: j=1, \ldots, N_{i,g}\}$ are i.i.d copies of a random variable $T$ with bounded density function within the domain $\mcT$. We are interested to test the null hypothesis 
\begin{equation}\label{eqn: OrigHypothesis}
\begin{aligned}
&H_0: \mu_1(t) = \mu_2(t) \;\; \forall \;\; t \in \mcT \\
   & \hspace{1.3 cm}\text{vs}  \\
   & H_1: \mu_1(t) \neq \mu_2(t) \;\; \text{for some} \;\; t \in \mcT.
    \end{aligned}
\end{equation}
We now provide a concise overview of the projection-based testing procedure by \cite{wang2021two}. Consider a sequence of orthonormal eigenfunctions, denoted as $\{\psi_k(t)\}_{k\geq 1}$, associated with the covariance operator $\Xi$, and ordered according to their eigenvalues: $\lambda_1 \geq \lambda_2 \geq \dots \geq 0$. Following the Kahrunen-Loeve theorem, the latent trajectories are projected onto the orthogonal eigenfunctions $\{\psi_k(t)\}_{k\geq 1}$ within the space $\mcL^2(\mcT)$, to represent the latent process as 
\begin{equation} \label{eqn: KLrep}
    X_{i,g}(t) = \mu_0(t) + \sum_{k=1}^\infty \zeta_{ik,g}\psi_{k}(t), \quad g=1, 2,
\end{equation}
where $\mu_0(t)$ is the pooled mean function common to both groups, and $\zeta_{ik,g} = \langle X_{i,g} - \mu_0, \psi_{k} \rangle$ is the group-specific projection that captures the group-specific difference. Specifically, $\{\zeta_{ik,g}\}$ are uncorrelated across different values of $k$, have a mean $\langle \mu_g-\mu_0, \psi_{k}\rangle$, and a variance $\lambda_{k}$ for $g=1,2$. To ensure identifiability, we set $\mu_g(t) - \mu_0(t) = 0$ for either $g=1$ or $g=2$. Until Section~\ref{sec: powerfn}, we assume that both the pooled mean $\mu_0(t)$ and the eigenfunctions are known. In this framework, the null hypothesis presented in (\ref{eqn: OrigHypothesis}) is equivalent to testing 
% \begin{equation*}
% \begin{aligned}
%    & H_0: \{\mathbb{E}(\zeta_{ik,1})\}_{k\geq 1} = \{\mathbb{E}(\zeta_{ik,2})\}_{k\geq 1} \\
%    & \hspace{2 cm} \text{v.s} \\
%    &H_1: \{\mathbb{E}(\zeta_{ik,1})\}_{k\geq 1} \neq  \{\mathbb{E}(\zeta_{ik,2})\}_{k\geq 1}.
%     \end{aligned}
% \end{equation*}
\begin{equation*}
\begin{aligned}
   & H_0: \{\mathbb{E}(\zeta_{ik,1})\}_{k\geq 1} = \{\mathbb{E}(\zeta_{ik,2})\}_{k\geq 1} \quad  \text{v.s} \quad H_1: \{\mathbb{E}(\zeta_{ik,1})\}_{k\geq 1} \neq  \{\mathbb{E}(\zeta_{ik,2})\}_{k\geq 1}.
    \end{aligned}
\end{equation*}

%See \cite{wang2021two} for detailed insights for the above statement. 
The projection-based test truncates the above hypothesis to the first $K$ dimensions and constructs a test-statistic using $\zeta_{ik,g}$, where $k=1, \dots, K$. The truncation parameter $K$ is usually chosen as the minimum number of eigenfunctions required to effectively explain the primary sources of variation in the data, often determined using metrics such as the Akaike Information Criterion (AIC), leave-one-out cross validation \citep{rice1991estimating}, or percentage of variation explained (PVE) \citep{xiao2016fast}. Under a sparse sampling design, the projections $\{\zeta_{ik,g}\}$ cannot be consistently estimated \citep{yao2005functional}. They are obtained via Best Linear Unbiased Prediction (BLUP) under the mixed model 
\begin{equation} \label{eqn: mixedmodel}
    \bY_{i,g} = {\boldsymbol{\mu}}_{0i,g} + \boldsymbol\Psi_{i,g} \boldsymbol{\zeta}_{i,g} +\boldsymbol{\epsilon}_{i,g},
\end{equation}
where $\bY_{i,g} = (Y_{i1,g}, \dots, Y_{iN_{i,g},g})^\top$ is the $N_{i,g}$-length stacked vector of response for the $i$th subject, ${\boldsymbol{\mu}}_{0i,g} = (\mu_0(T_{i1,g}), \dots, \mu_0(T_{iN_{i,g},g}))^\top$, and $\boldsymbol\Psi_{i,g} =(\bpsi_{i1,g}, \dots, \bpsi_{iK,g})$ is the column-stacked version of $\{\bpsi_{ik,g}\}_{k=1}^K$'s with ${\boldsymbol{\psi}}_{ik,g} = (\psi_k(T_{i1,g}), \dots, \psi_k(T_{iN_{i,g},g}))^\top$ is the mean vector and the eigenvector matrix, ${\boldsymbol\zeta}_{i,g} = ({\zeta}_{i1,g}, \dots, {\zeta}_{iK, g})^\top$ is the vector of projections, and $\boldsymbol\epsilon_{i,g}$ is the vector of measurement errors for the $i$th subject. Under a `working' Gaussian assumption of the scores, the BLUP of $\boldsymbol\zeta_{i,g}$ under model~(\ref{eqn: mixedmodel}) is of the form
\begin{equation} \label{eqn: BLUP}
\begin{aligned}
    \widetilde{\boldsymbol\zeta}_{i,g} := \mathbb{E}(\boldsymbol\zeta_{i,g} \vert \bY_{i,g}) = \text{diag}(\lambda_1, \dots, \lambda_K) {\boldsymbol\Psi}_{i,g}^{\top} \bG_{\bY_{i,g}}^{-1} (\bY_{i,g} - \boldsymbol\mu_{0i,g}),
    \end{aligned}
\end{equation}
where $\bG_{\bY_{i,g}} = \text{Cov}(\bY_{i,g}) = \{\Sigma(T_{ij,g}, T_{ij^\prime,g}) + \tau^2 \mathbb{I}(j = j^\prime)\}_{1 \leq j, j^\prime \leq N_{i,g}}$ is the covariance matrix of $\bY_{i,g}$. The quantity $\widetilde{\boldsymbol\zeta}_{i,g}$ is termed as `shrinkage' scores, and it is a rational choice for estimating the true score $\boldsymbol\zeta_{i,g}$ because $\mathbb{E}(\widetilde{\boldsymbol\zeta}_{i,g}) = \mathbb{E}[\mathbb{E}(\boldsymbol\zeta_{i,g} \vert \bY_{i,g})] = \mathbb{E}(\boldsymbol\zeta_{i,g})$. The shrinkage scores $\widetilde{\boldsymbol\zeta}_{i,g}$ are a consistent estimator of the unobserved scores $\boldsymbol\zeta_{i,g}$ as the number of observations per subject grows and the measurement error gets small. The population covariance of the `shrinkage' scores can be expressed as
\begin{align}
    \nonumber \boldsymbol\Lambda_g = \text{Cov}(\widetilde{\boldsymbol\zeta}_{i,g}) &= \mathbb{E}_T\,\left[\text{diag}(\lambda_1, \dots, \lambda_K) {\boldsymbol\Psi}_{i,g}^{\top} \bG_{\bY_{i,g}}^{-1} {\boldsymbol\Psi}_{i,g}\text{diag}(\lambda_1, \dots, \lambda_K)\right] \\
    & \qquad + \text{Cov}_{T}\left[\text{diag}(\lambda_1, \dots, \lambda_K) {\boldsymbol\Psi}_{i,g}^{\top} \bG_{\bY_{i,g}}^{-1} (\mathbb{E}(\mathbf{Y}_{i,g}) - \boldsymbol\mu_{0i,g})\right], \label{eqn: cov_shrinkage}
\end{align}
where $\mathbb{E}_T$ and $\text{Cov}_T$ denotes the expectation and covariance under the sampling distribution of the design points. From the above expression, it is evident that $\{\widetilde{\zeta}_{ik,g}\}_{k=1}^K$ are  not uncorrelated across $k$, unlike the true scores ${\zeta}_{ik,g}$.

These `shrinkage' scores are the building blocks of the projection-based test. Under the null hypothesis, the `shrinkage' score enjoys some nice properties. Under the $H_0$, $\mathbb{E}(\widetilde{\boldsymbol\zeta}_{i,1} - \widetilde{\boldsymbol\zeta}_{i,2}) = 0$, and $\boldsymbol\Lambda_1 = \boldsymbol\Lambda_2$  since the last term in~(\ref{eqn: cov_shrinkage}) is zero. See Eq. (2.11) of \cite{wang2021two} for more details on this point. \cite{wang2021two} used these properties of the `shrinkage' scores to construct a statistic for testing $H_0$. We define $\widetilde{\boldsymbol\zeta}_{g+} = n_g^{-1}\sum_{i = 1}^{n_g} \widetilde{\boldsymbol\zeta}_{i,g}$ and $\widetilde{\boldsymbol\Lambda}_g = (n_g-1)^{-1}\sum_{i= 1}^{n_g} (\widetilde{\boldsymbol\zeta}_{i,g} - \widetilde{\boldsymbol\zeta}_{g+})(\widetilde{\boldsymbol\zeta}_{i,g} - \widetilde{\boldsymbol\zeta}_{g+})^\top$ as the group-specific average  and the sample covariance of the `shrinkage' scores. Further, we define the pooled sample covariance as $\widetilde{\boldsymbol\Lambda} = \{(n_1-1)\widetilde{\boldsymbol\Lambda}_1 + (n_2-1)\widetilde{\boldsymbol\Lambda}_2\}/(n_1 + n_2-2)$. To test for $H_0$, a Hotelling $T^2$ random variable \citep{hotelling1992generalization} using the `shrinkage' scores can be constructed as
   \begin{equation} \label{eqn: HotellingT2}
     T_{n}  = \frac{n_1n_2}{n_1+n_2} (\widetilde{\boldsymbol\zeta}_{1+} - \widetilde{\boldsymbol\zeta}_{2+})^\top \widetilde{\boldsymbol\Lambda}^{-1} (\widetilde{\boldsymbol\zeta}_{1+} - \widetilde{\boldsymbol\zeta}_{2+}). 
\end{equation}
The test rejects $H_0$ at a specified significance level $\alpha \in (0,1)$ if 
\begin{equation} \label{eqn: HotellingTestRule_eqvar}
    {T}_n  > \frac{(n-2)K}{n - K - 1}F_\alpha( K ,n - K - 1),
\end{equation}
where $K$ is the dimension of `shrinkage' scores estimated from the data, $n=n_1 + n_2$ is the total number of subjects, and $F_\alpha(a,b)$ is the $100(1-\alpha)\%$ quantile of $F$-distribution with $a$ and $b$ degrees of freedom.

\subsection{Exact non-null distribution of the test}
We now present our major contributions. In the next two sections, we will derive the alternate distribution of the test statistic, which is pivotal for constructing the PASS formula detailed in Section~\ref{sec: PASS}. We assume that the data follow a sparse design in line with the underlying model~(\ref{eqn: dgm}). In the subsequent proposition, we present the distribution of $T_n$ under scenarios where the group difference $\mu_1(t) - \mu_2(t)$ is non-zero and the eigenfunctions are known.

The construction of Hotelling $T^2$ statistic, using the pooled covariance $\widetilde{\boldsymbol\Lambda}$ in~(\ref{eqn: HotellingT2}), relies on an essential observation --- the population covariances of the `shrinkage' scores, $\boldsymbol{\Lambda}_g$ in~(\ref{eqn: cov_shrinkage}), are equal for both groups, a distinctive feature that is satisfied only under the null hypothesis. However, under the alternative hypothesis, these covariances differ due to the non-null disparity in mean functions between the groups. This heteroscedasticity poses a significant challenge in deriving the non-null distribution of the test statistic. To accurately compute the power function of the projection-based test, it becomes crucial to derive the non-trivial alternate distribution of the Hotelling $T^2$ statistic under unequal variances of the `shrinkage' scores between the two groups. The form of the non-null distribution of the Hotelling $T^2$ statistic is presented in the next theorem under fixed sample size.

\begin{thm} \label{thm: power_egnknown}
Assume that model~(\ref{eqn: dgm}) for the observed response $\{\bY_{ij,g}: j=1, \dots, N_{i,g}\}_{i=1}^{n_g}$ is true, and $\sup_{i,g} N_{i,g} < \infty$ almost surely. Assume that the scores $\{\zeta_{ik,g}\}$ are independent across $i$ and $g$ and they are Gaussian. Further assume that the true mean functions $\mu_g(t)$ and the true eigencomponents $\{\lambda_k, \psi_k(t)\}_{k\geq 1}$ are known. Then, conditional on the truncation parameter $K$,
%$$
%\frac{(n - K - 1)T_n}{(n-2)K} \sim F(K,  n - K - 1, \frac{n_1n_2}{n_1 + n_2} \boldsymbol{\Delta}^\top \boldsymbol\Lambda^{-1}\boldsymbol{\Delta}),
%$$
\begin{align}
\frac{n_2(1 + 1/\kappa)T_n}{n_1 + n_2 - 2}  \;\;\overset{d}{=} \;\;\frac{\sum_{k=1}^K d_k^{-1} \chi^2_1 \left (n_1(\boldsymbol{u}_k^\top {\boldsymbol\Lambda^{\dagger}}^{-1/2} \boldsymbol\Delta)^2  \right )}{ \chi^2_{\nu - K + 1}/\nu} \label{eqn: alt_distr_ht}
\end{align}
where $\boldsymbol{\Delta} = (\delta_1, \dots, \delta_{K})^\top$ with $\delta_k =  \int_{\mcT} \{\mu_1(t) - \mu_2(t) \}\psi_k(t) dt$ and $\boldsymbol\Lambda^{\dagger} = \boldsymbol\Lambda_1 + \kappa\boldsymbol\Lambda_2$, with ${\boldsymbol\Lambda}_g : g=1,2$ are the population covariance of the `shrinkage'  scores as in~(\ref{eqn: cov_shrinkage}),  and $\kappa = n_1/n_2$, is the allocation ratio of the samples in the group.  Moreover, $\{d_k, \boldsymbol{u}_k\}_{k=1}^K$ are the eigenvalues and the eigenvectors in the spectral decomposition of ${\boldsymbol\Omega^\dagger} = \sum_{k=1}^K d_k \boldsymbol{u}_k\boldsymbol{u}_k^\top$, with $\boldsymbol\Omega^\dagger : =  \kappa (\kappa - 1/n_2)  \boldsymbol\Omega + (1-1/n_2)(\mathbf{I}_K -  \boldsymbol\Omega)$, with $ \boldsymbol\Omega : = {\boldsymbol\Lambda^{\dagger}}^{-1/2}\boldsymbol\Lambda_1{\boldsymbol\Lambda^{\dagger}}^{-1/2}$. Finally, $\nu$ is the degrees of freedom of the chi-squared distribution in the denominator with the form, 
{\small\begin{align*}
\nonumber \nu &= n_2 \left\{\tr({\boldsymbol\Omega^\dagger}^2) + [\tr(\boldsymbol\Omega^\dagger)]^2\right\}  \left[\kappa^2(\kappa-n_2^{-1})\{\tr({\boldsymbol\Omega}^2)  \right. \\
&  \qquad \;\;+ [\tr(\boldsymbol\Omega)]^2\} + \left. (1-n_2^{-1})\{\tr[({\mathbf{I}_K - {\boldsymbol\Omega}})^2] + [\tr(\mathbf{I}_K - \boldsymbol\Omega)]^2\} \right]^{-1},
\end{align*}
}
where $\tr(\bA)$ means the sum of diagonal elements of a matrix $\bA$, and for two random variables $U$ and $V$, the notation $U \overset{d}{=} V$ means the distribution of $U$ and $V$ are same.
\end{thm}
Theorem~\ref{thm: power_egnknown} gives the exact distribution of the test-statistic $T_n$ under the assumption of known eigenfunctions and Gaussian-distributed scores. While this distribution does not precisely follow an $F$ distribution, it closely resembles a non-central $F$ distribution. The numerator is a linear combination of $K$ independent non-central chi-squared random variables, each with one degree of freedom, and the denominator is a $\chi^2_{\nu - K +1}$. The proof of Theorem~\ref{thm: power_egnknown} is presented in Section~\ref{sec: proof_of_theorems} of Supplementary material, and involves deriving the alternate distribution for the Hotelling $T^2$ statistic when data have unequal variances across two groups. The degree of freedom parameter $\nu$ is complicated and it originates from the approximation of the sum of two Wishart random variables to another Wishart random variable \citep{nel1986solution}. Although the non-null distribution is complex, we can efficiently simulate random samples from it to calculate the test's power function. In the subsequent section, we expand on the statement of the Theorem~\ref{thm: power_egnknown} when eigenfunctions are unknown and provide insights into practical methods for calculating the test's power function.

% \href{https://youtu.be/75juXt8P_bg}

\subsection{Asymptotic non-null distribution of the test}  \label{sec: powerfn}
In practical scenarios, both the mean $\mu_0(t)$ and the eigenfunctions $\{\psi_k(t)\}_{k=1}^K$ are unknown.  These are estimated through functional principal component analysis (fPCA) \citep{yao2005functional} applied to the data. For both dense and sparse functional designs, the estimated mean and eigenfunctions are consistent as the number of subjects $n$ approaches infinity \citep{zhang2016sparse}. Further details on the estimation of the mean and eigenfunctions can be found in Section 2.3 of \cite{wang2021two}. When plugging in the estimator for the common mean $\mu_0(t)$ and $\{\psi_k(t)\}_{k=1}^K$ into (\ref{eqn: BLUP}), we obtain $\widehat{\boldsymbol\zeta}_{i,g} = (\widehat{\zeta}_{i1,g}, \dots, \widehat{\zeta}_{iK, g})^\top$ as the estimated `shrinkage' scores for the $i$th subject. Here, $\widehat{\boldsymbol\zeta}_{1+} = n_1^{-1}\sum_{i = 1}^{n_1} \widehat{\boldsymbol\zeta}_{i,1}$ and the $\widehat{\boldsymbol\zeta}_{2+} = n_2^{-1}\sum_{i=1}^{n_2} \widehat{\boldsymbol\zeta}_{i,2}$ represent the group-specific averages of the estimated scores. Similarly, we define $\widehat{\boldsymbol\Lambda}_1 = (n_1-1)^{-1}\sum_{i= 1}^{n_1} (\widehat{\boldsymbol\zeta}_{i,1} - \widehat{\boldsymbol\zeta}_{1+})(\widehat{\boldsymbol\zeta}_{i,1} - \widehat{\boldsymbol\zeta}_{1+})^\top$ and $\widehat{\boldsymbol\Lambda}_2 = (n_2-1)^{-1}\sum_{i = 2}^{n_2} (\widehat{\boldsymbol\zeta}_{i,2} - \widehat{\boldsymbol\zeta}_{2+})(\widehat{\boldsymbol\zeta}_{i,2} - \widehat{\boldsymbol\zeta}_{2+})^\top$ as the sample variances. We define $\widehat{\boldsymbol\Lambda} = \{(n_1-1)\widehat{\boldsymbol\Lambda}_1 + (n_2-1)\widehat{\boldsymbol\Lambda}_2\}/(n-2)$ as the pooled sample covariance. In practical situations, the  null hypothesis $H_0$ is tested based on the statistic  
\begin{equation*} %\label{eqn: HotellingT2_prac}
     \widehat{T}_{n}  = \frac{n_1n_2}{n_1+n_2} (\widehat{\boldsymbol\zeta}_{1+} - \widehat{\boldsymbol\zeta}_{2+})^\top \widehat{\boldsymbol\Lambda}^{-1} (\widehat{\boldsymbol\zeta}_{1+} - \widehat{\boldsymbol\zeta}_{2+}).
\end{equation*}
It is worth noting that the test statistic $\widehat{T}_n$ represents the observed counterpart of the unobserved quantity $T_n$ defined in~(\ref{eqn: HotellingT2}). \cite{wang2021two} established that the asymptotic null distribution of the test statistic $\widehat{T}_n$ follows a $\chi^2$ distribution with $K$ degrees of freedom. In the next proposition, we will derive the asymptotic non-null distribution of the test-statistic $\widehat{T}_n$, assuming that eigenfunctions are consistently estimated from the data.

\begin{prop} \label{thm: power_egnunknown}
Assume that model~(\ref{eqn: dgm}) for the observed response $\{\bY_{ij,g}: j=1, \dots, N_{i,g}\}_{i=1}^{n_g}$ is true, and $\sup_{i,g} N_{i,g} < \infty$ almost surely. Suppose that the alternate hypothesis is true and that the true mean difference is characterized by $$H_1: \mu_1(t) - \mu_2(t) = n^{-\varrho/2}\eta_0(t),$$ with $\eta_0(t) \ne 0$, is a fixed known function of $t$, for some $\varrho \in [0,1]$. Assume that the mean functions for both the groups and the covariance components are estimated consistently, i.e., $\lVert \widehat{\bmu}_g - \bmu_g\rVert = o_p(1)$ for both $g=1,2$, $\lVert \widehat{\boldsymbol\psi}_k - \boldsymbol\psi_k \rVert = o_p(1)$, $\lVert \widehat{\lambda}_k - \lambda_k\rVert = o_p(1)$ for all $k=1, \dots, K$, and  $\lvert \widehat{\tau}^2 - \tau^2\rvert = o_p(1)$, and that $\lim n_1/n_2 \to \kappa \in (0, \infty)$. Then, conditional on the truncation parameter $K$,
%$$
%\widehat{T}_n \overset{d}{\to} \chi^2_K(\boldsymbol{\Delta}^\top \boldsymbol\Lambda^{-1}\boldsymbol{\Delta}),
%$$
\begin{equation*}
\widehat{T}_n \overset{d}{\longrightarrow} \begin{cases}  \kappa {\sum_{k=1}^K d_k^{-1} \chi^2_1 \left ((\boldsymbol{u}_k^\top {\boldsymbol\Lambda^{\dagger}}^{-1/2} \boldsymbol\Delta)^2  \right )} &\quad \text{if } \varrho = 1 \\
\infty &\quad \text{if } \varrho < 1
\end{cases}
\end{equation*}
where $\boldsymbol{\Delta} = (\delta_1, \dots, \delta_{K})^\top$ with $\delta_k = \langle \eta_0, \psi_k \rangle$ and $\boldsymbol\Lambda^{\dagger} = \boldsymbol\Lambda_1 + \kappa\boldsymbol\Lambda_2$, with $\{{\boldsymbol\Lambda}_g : g=1,2\}$ are the population covariance of the `shrinkage'  scores, defined in (\ref{eqn: cov_shrinkage}). Moreover, $\{d_k, \boldsymbol{u}_k\}_{k=1}^K$ are the eigenvalue and the eigenvectors of the spectral decomposition ${\boldsymbol\Omega^\dagger} = \sum_{k=1}^K d_k \boldsymbol{u}_k\boldsymbol{u}_k^\top$, with $\boldsymbol\Omega^\dagger : =  \mathbf{I}_K + (\kappa^2-1) \,{\boldsymbol\Lambda^{\dagger}}^{-1/2}\boldsymbol\Lambda_1{\boldsymbol\Lambda^{\dagger}}^{-1/2}.$
\end{prop}
Proposition~\ref{thm: power_egnunknown} presents the asymptotic distribution of the projection-based test under the local alternative hypothesis. Specifically, when the true group difference decays at a rate slower than $n^{-1/2}$, the test rejects the null hypothesis with probability $1$. It is noteworthy that \cite{wang2021two} demonstrates the convergence of the test's power function to 1, but the author does not provide a specific form for the alternate distribution of the test statistic. We remark that the alternate distribution does not require a Gaussian assumption for the `shrinkage' scores, and Proposition~\ref{thm: power_egnunknown} relies on the consistent estimation of eigenfunctions from the observed data.

\section{Power and sample size (PASS) calculation formula}\label{sec: PASS}

\subsection{Power function} \label{sec: power}
To investigate test's power across various sample sizes and determine the minimum required sample size (discussed in Section \ref{sec: sampsize}), we use the non-null distribution of the test-statistic as in Theorem~\ref{thm: power_egnknown}, under a working Gaussian distributional assumption of the `shrinkage' scores. Let $\eta(t) := \mu_1(t)-\mu_2(t)$ be the pre-specified mean difference function between the two groups. For simplicity, we will subsequently use $\eta(t)$ to represent the group mean difference. For eigenfunctions $\{\psi_k(t)\}_{k=1}^K$, and $(n_1, n_2)$ being the number of subjects in the groups, let $\mathcal{F}(x)$ be the cumulative distribution function (CDF) of the random variable in the right side of~(\ref{eqn: alt_distr_ht}) obtained by replacing $\mu_1(t)-\mu_2(t)$ with mean difference function $\eta(t)$ in Theorem~\ref{thm: power_egnknown}. The theoretical power function of the test can be expressed as  
\begin{equation} \label{eqn: powerfn}
    \mathcal{P}_{n, \kappa}[\eta(t)] = 1 - \mathcal{F}\left(\frac{K n_2 F_\alpha( K ,n - K - 1) }{ (1 + 1/\kappa)^{-1}(n-K-1)}\right). 
 %\mathbb{P}\left(F^* > \frac{K n_2 F_\alpha( K ,n - K - 1) }{ (1 + 1/\kappa)^{-1}(n-K-1)} \right)
\end{equation}
The notation $\mathcal{P}_{n,\kappa}[\eta(t)]$ denotes the power function of the projection-based test for total sample size $n = n_1 + n_2$, the allocation ratio $\kappa = n_1/n_2$, and mean difference function $\eta(t)$. For a given pair of parameters $n$ and $\kappa$, we can determine the group sizes as $n_1 = n\kappa/(1+\kappa)$ and $n_2 = n/(1+\kappa)$. Eqn.~(\ref{eqn: powerfn}) provides the working formula for the test's theoretical power. To obtain an accurate empirical estimate of the CDF $\mathcal{F}(x)$, we can generate a large number of random samples from the distribution described on the right-hand side of~(\ref{eqn: alt_distr_ht}). From a practical standpoint, when the mean difference $\eta(t)$ and the eigenfunctions $\{\psi_k(t)\}_{k=1}^K$ are hypothesized, one can compute the power function for any $(n_1, n_2)$ provided that we have some knowledge about $\boldsymbol\Lambda_1$ and $\boldsymbol\Lambda_2$, the unknown covariance parameters of the `shrinkage` scores for both groups. Notably, these covariance parameters, as perceived from Eq. (\ref{eqn: cov_shrinkage}), are complicated functions of the mean, eigenfunctions of the stochastic process, and the sampling distribution of observation points. In the next section, we will introduce an efficient algorithm for computing $\mathcal{P}_{n, \kappa}(\eta)$ by reliably estimating the unknown covariances $\{\boldsymbol\Lambda_g\}_{g=1}^2$ from a representative large sample, without deriving it's complex mathemtical form.
 
\subsubsection{Data-driven estimation of theoretical power function} \label{sec: datadrivepower}
The power function formula in~(\ref{eqn: powerfn}) requires prior knowledge of the true eigenfunctions, limiting its practical use. In real-world scenarios, understanding the covariance structure $\Sigma(t,t^\prime)$ of the observational process, rather than having explicit knowledge about the eigenfunctions, is more practical. Therefore, a method for extracting eigenfunctions from the covariance function $\Sigma(t,t^\prime)$ is essential for power function computation. 

When the covariance process $\Sigma(t,t^\prime)$ is specified, consistent estimates of the eigenfunctions can be obtained via spectral decomposition of the covariance matrix evaluated a fine grid of points. Specifically, evaluating the $R \times R$ covariance matrix $\boldsymbol\Sigma = (\Sigma(t_{r}, t_{r^\prime})_{1 \leq r , r^\prime \leq R}$ at a fine grid $t_1 < t_2 < \ldots < t_R$ of length $R$ in the domain $\mathcal{T}$, with $R$ being substantially large, we can perform a spectral decomposition of $\boldsymbol\Sigma$. This process yields reliable estimates of the true eigenfunctions $\{\widetilde{\psi}_k(t) : t=t_1, \dots, t_R\}_{k=1}^{K}$. The number of leading eigenfunctions, ${K}$, can be determined based on a pre-specified percentage of variation explained (PVE), often set at a number higher than 90\%. Next, the projection of the mean difference on the eigenfunctions can be numerically computed as $\delta_k = R^{-1}\sum_{r=1}^R \eta(t_r)\widetilde{\psi}_k(t_r)$ for $k=1, \dots, K$. However, to compute the power function in~(\ref{eqn: powerfn}), we still need a reliable estimate of the unknown covariances of the `shrinkage' scores, $\{\boldsymbol\Lambda_g : g=1,2\}$. 

To obtain consistent estimators for both the eigenfunctions and the covariance of the `shrinkage' scores in a unified manner, we propose a data-driven approach. Specifically, for a given mean difference function $\eta(t)$ and covariance $\Sigma(t,t^\prime)$, we generate synthetic data with a potentially large number of subjects, e.g., 10000, equally allocated to each group. We simulate this data using model~(\ref{eqn: dgm}) with mean functions $\mu_1(t) = 0$ and $\mu_2(t) = \eta(t)$, and covariance specified by $\Sigma(t,t^\prime)$. The data for each subject are generated based on a pre-specified number of observations and schedules. Then, we employ standard statistical software tools designed for fPCA under sparse designs, such as R package \verb|face| \citep{face}. This software internally estimates the eigenfunctions and obtains highly reliable estimates of $\boldsymbol \Lambda_g = \text{Cov}(\widetilde{\boldsymbol\zeta}_{i,g})$ by computing the sample covariance of `shrinkage' scores based on the large number of subjects allocated in each group. Since these quantities are estimated from a large sample size, they are expected to closely approximate the true eigenfunctions and true covariance of the `shrinkage' scores. The computation details are specified in Algorithm~\ref{alg: powerfunction}. 

\begin{algorithm}
 \SetAlgoLined
 \KwIn{Data information: True difference in mean function between two groups, i.e., $\eta(t) := \mu_1(t) - \mu_2(t)$, covariance of the latent process $\Sigma(t,t^\prime)$, measurement error variance $\tau^2$, significance level $\alpha \in (0,1)$, sample sizes $(n_1, n_2)$, and a pre-specified PVE to estimate the eigencomponents of the covariance\;}
 Generate a dataset with large number of subjects, say $5000$, using  the model~(\ref{eqn: dgm}) by setting $\mu_{1}(t) = 0$ and $\mu_{2}(t) = \eta(t)$ and covariance of the $X_g(t)$ as $\Sigma(t,t^\prime)$ and the measurement error variance $\tau^2 > 0$. \;
 Preset a large grid of size $R$ (say 100) and generate a sequence of points $t_1 < t_2, \ldots < t_R$ in $\mcT$ \;
 Conduct fPCA on the generated dataset to obtain a highly reliable estimate of the eigenfunctions $\{\widetilde{\psi}_k(t) : k=1, \dots, K\}$ at $t=t_1, \dots, t_R$. The number of eigenfunctions $K$ are chosen by the specified PVE \;
 Calculate the fPC scores for each subject and obtain the sample covariances $\{\widetilde{\boldsymbol\Lambda}_g : g=1,2\}$ of the scores for the two groups. This will serve as a reliable estimated of the true covariances $\{\boldsymbol\Lambda_g : g=1,2\}$. \;
 		\For{$k \in \{1, \ldots, K\}$}{
 		 Calculate the projection $\widetilde{\delta}_k = \int \eta(t)\widetilde{\psi}_k(t)dt \approx R^{-1}\sum_{r=1}^R \eta(t_r)\widetilde{\psi}_k(t_r)$
			 
 		}
Construct the vector $\widetilde{\boldsymbol{\Delta}} = (\widetilde{\delta}_1, \dots, \widetilde{\delta}_{{K}})^\top$\;
 		Return power function $\mathcal{P}_{n,\kappa}[\eta(t)]$ as in Equation~(\ref{eqn: powerfn}) by replacing $\boldsymbol{\Delta}$ and $\{\boldsymbol{\Lambda}_g\}_{g=1}^2$ with $\widetilde{\boldsymbol{\Delta}}$ and $\{\widetilde{\boldsymbol{\Lambda}}_g\}_{g=1}^2$ respectively\; 
 		 \caption{Algorithm for power function $\mathcal{P}_{n, \kappa}(\eta(t))$ of the projection-based test.}
 		  \label{alg: powerfunction}
 \end{algorithm}

Steps 1-4 of Algorithm~\ref{alg: powerfunction} are conducted only once to reliably estimate the true eigenfunctions and covariance parameter of the `shrinkage' scores, a critical requirement for accurate power estimation as mentioned in Proposition~\ref{thm: power_egnunknown}. Once these quantities are consistently estimated, we apply the formula in~(\ref{eqn: powerfn}) substituting the true quantities with their estimated counterparts to reliably compute the test's power function. In the longitudinal data analysis, common covariance assumptions are often stationary, such as compound symmetric or autoregressive (AR) structures. However, Algorithm~\ref{alg: powerfunction} can compute the test's power function for any general covariance structure of the stochastic process $X_g(t)$.

Two important remarks conclude this section. First, the power function in~(\ref{eqn: powerfn}) and the empirically calculated power (based on a large number of replications) closely align only if the sample size $n$ is sufficiently large for consistent eigenfunction estimation. Empirical power refers to conducting the test~(\ref{eqn: HotellingTestRule_eqvar}) on a large number, say $B$, of datasets (e.g., $B=1000$) generated using the specified mean difference and covariance structure. The recorded percentage of times the test rejects $H_0$ provides empirical power. If the sample size is too small for consistent eigenfunction estimation, the asymptotic null distribution of the test statistic $\widehat{T}_n$ no longer follows a chi-squared distribution with $K$ degrees of freedom. Consequently, empirically computed power using test-rule~(\ref{eqn: HotellingTestRule_eqvar}) based on $B$ datasets becomes unreliable and should not be compared to that obtained from~(\ref{eqn: powerfn}). Second, the test's power function is minimally affected by an increased percentage of missing observations per subject (up to a certain level). This robustness arises because the quality of estimating $\{\widehat{\psi}_k(t)\}_{k=1}^K$ does not deteriorate with fewer number of observations per subject, provided the eigenfunctions are estimated from a large number of subjects. See \cite{zhang2016sparse} for theoretical results on the consistency of eigenfunctions when the number of observations subjects is fixed, under a sparse design. Consequently, for large sample sizes, missing data has limited impact, rendering the test 'missing-immune.' This phenomenon is numerically demonstrated in Section~\ref{sec: sim_missing_immune}. 
 
\subsection{Sample size calculation} \label{sec: sampsize}
In this section we introduce an algorithm for calculating the minimum sample size required for the test to achieve a  target power level of $100\gamma\%$ for any $\gamma \in (0,1)$ to detect a non-null group difference specified by $\eta(t): = \mu_1(t) - \mu_2(t)$, based on the power function formula provided in~(\ref{eqn: powerfn}).
 
 \begin{prop} \label{thm: sampsize_egnknown}
 Assume that the conditions of the Theorem~\ref{thm: power_egnknown} is true. Suppose that the difference of the mean function between the two groups are specified by $\eta(t)$. Assume that the eigenfunctions of the covariance of $X_g(t)$ is given, so that we have the projection-vector $\boldsymbol{\Delta} = (\delta_1, \dots, \delta_{K})^\top$ with $\delta_k = \int\eta(t) \psi_k(t)\,dt $ and the population covariance of the `shrinkage' scores $\{\boldsymbol\Lambda_g : g=1,2\}$ is known. Let $\kappa$ be pre-specified allocation ratio of the samples in the group. Then the minimum sample size required for the test to achieve a power of $100\gamma\%$ is $(\kappa n^*, n^*)$ where $n^*$ is the minimum positive integer that satisfies
{\small\begin{align}
\nonumber \mathbb{P}\, &\left(\frac{\sum_{k=1}^K d_k^{-1} \chi^2_1 \left (\kappa n^* (\boldsymbol{u}_k^\top {\boldsymbol\Lambda^{\dagger}}^{-1/2} \boldsymbol\Delta)^2  \right )}{ \chi^2_{\nu(n^*) - K + 1}/\nu(n^*)} \right. \\
&\qquad  \qquad > \left. \frac{K  n^* F_\alpha( K ,  (1+\kappa)n^* - K - 1) }{ (1 + 1/\kappa)^{-1}( (1+\kappa)n^*-K-1)} \right) > \gamma,  \label{eqn: samplesize_fixed}
\end{align}}
\noindent where $\boldsymbol\Lambda^{\dagger} = \boldsymbol\Lambda_1 + \kappa\boldsymbol\Lambda_2$, and $\{d_k, \boldsymbol{u}_k\}_{k=1}^K$ are the eigenvalues and the eigenvectors in  the spectral decomposition of ${\boldsymbol\Omega^\dagger} = \sum_{k=1}^K d_k \boldsymbol{u}_k\boldsymbol{u}_k^\top$, with $\boldsymbol\Omega^\dagger : =  \kappa (\kappa - 1/n^*)  \boldsymbol\Omega + (1-1/n^*)(\mathbf{I}_K -  \boldsymbol\Omega)$, with $ \boldsymbol\Omega : = {\boldsymbol\Lambda^{\dagger}}^{-1/2}\boldsymbol\Lambda_1{\boldsymbol\Lambda^{\dagger}}^{-1/2}$. Finally, $\nu(n^*)$ is the degrees of freedom of the chi-squared distribution which is of the form
{\small\begin{align*}
\nonumber &\nu(n^*) = n^* \left\{\tr({\boldsymbol\Omega^\dagger}^2) + \tr^2(\boldsymbol\Omega^\dagger)\right\}  \left[\kappa^2(\kappa-{1/n^*})\{\tr({\boldsymbol\Omega}^2)  \right. \\
& \qquad \qquad \qquad   + \tr^2(\boldsymbol\Omega)\} + \left. (1- 1/n^*)\{\tr({\mathbf{I}_K - {\boldsymbol\Omega}})^2 + \tr^2(\mathbf{I}_K - \boldsymbol\Omega)\} \right]^{-1}.
\end{align*}
}
%\begin{multline} \label{eqn: samplesize_fixed}
%\mathbb{P}\,(F(K,  (1+\kappa)n^* - K - 1, n^* (1+\kappa^{-1})^{-1}\boldsymbol{\Delta}^\top \boldsymbol\Lambda^{-1}\boldsymbol{\Delta})  \\
%   >  F_\alpha( K , (1+\kappa)n^* - K - 1) ) > \gamma. 
%\end{multline} 
\end{prop}

The minimum value of $n^*$ appears on both sides of the equation above. It can be determined using a `brute-force' approach, by iteratively calculating the left-hand side for various $n^*$ values, starting from an initial value of $3$ and stopping when it exceeds $\gamma$. Alternatively, standard root-solver algorithms in software (e.g., \verb|uniroot()| function in R) can be used to find the $n^*$ value that makes both sides equal. In practical scenarios, where the true eigenfunctions and covariances of `shrinkage' scores are unknown, we employ the data-driven approach from Section~\ref{sec: datadrivepower} to reliably estimate these unknown quantities for sample size calculation, using formula~(\ref{eqn: samplesize_fixed}). Specifically, we reuse Steps 1-8 of Algorithm~\ref{alg: powerfunction} to create an efficient algorithm for determining sample sizes for our projection-based test, outlined in Algorithm~\ref{alg: sampsize}.

\begin{algorithm}
 \SetAlgoLined
 \KwIn{Data information: True difference in mean function between two groups, i.e., $\eta(t) := \mu_1(t) - \mu_2(t)$, covariance of the latent process $\Sigma(t,t^\prime)$, measurement error variance $\tau^2$, significance level $\alpha \in (0,1)$, target power $\gamma \in (0,1)$, ratio of the sample size between two groups, $\kappa = n_1/n_2 > 0$, and a pre-specified PVE to estimate the eigencomponents of the covariance\;}
Repeat Step 1-8 of Algorithm~\ref{alg: powerfunction} \;
 	Replace  $\boldsymbol{\Delta}$ and $\{\boldsymbol{\Lambda}\}_{g=1}^2$  with $\widetilde{\boldsymbol{\Delta}}$ and $\{\widetilde{\boldsymbol{\Lambda}}\}_{g=1}^2$ respectively in~(\ref{eqn: samplesize_fixed}) to calculate the minimum value of $n^*$ such that~(\ref{eqn: samplesize_fixed}) holds \; 
	Return the minimum sample size required for the two groups as $(\kappa n^*, n^*)$. 
 		 \caption{Sample size calculation algorithm for projection-based test.}
 		  \label{alg: sampsize}
 \end{algorithm}

Sample size calculation Algorithm~\ref{alg: sampsize} requires the true group difference, the covariance function, observation schedule, and group allocation ratio to compute the minimum sample size for achieving a $100\gamma\%$ power level. The most time-consuming part involves the first four steps, which require generating extensive data and conducting fPCA to compute eigenfunctions and `shrinkage' score covariance. This initial task is a one-time requirement and cannot be avoided, as consistent estimates of $\{\psi_k(t)\}_{k=1}^K$ and $\{\boldsymbol\Lambda_g\}_{g=1}^2$ are crucial for determining an appropriate sample size. Subsequent steps in the algorithm require minimal computing time once these estimates are available.

\subsection{Computational software}
A user-friendly R package \verb|fPASS| \citep{fPASS2023} implementing the Algorithm~\ref{alg: powerfunction} and~\ref{alg: sampsize} is available at CRAN, \url{https://cran.r-project.org/web/packages/fPASS/index.html}. Its capabilities include (1) versatility in handling stationary and non-stationary covariance structures, (2) flexible visit scheduling options (fixed or random) for individual subjects, (3) user-friendly tools for specifying missing data percentages, (4) adaptability to various forms of mean difference functions between groups. The package's comprehensive functionality covers all options available in the GEE module of NCSS PASS software \citep{PASS}. Notably, it efficiently computes power functions and sample sizes for both longitudinal and functional designs. Detailed instructions and case study examples are available in the package vignette for a more comprehensive understanding of its application.

\section{Numerical studies} \label{sec: edfd_testing_simstudy}
\subsection{Data generation} \label{sec: edfd_testing_simstudy:dataGen}
In this section, we conduct empirical validations to assess the test's power function across varying sample sizes and verify the accuracy of our sample size calculations. We treat the sample size calculations as an inverse problem originating from the power function validation. To achieve this, we perform a series of focused numerical experiments that demonstrate the validity of the power function formula. These experiments involve three key factors. We take the difference in the mean function between the two groups as $\mu_1(t) - \mu_2(t) = \eta t^3$, with the scalar variable $\eta$ taking different values greater than zero to ensure deviation from the null hypothesis. We consider three different (common) covariance structures of the stochastic process $X_g(t)$, with the first two being stationary while the last one being non-stationary. \\
\noindent Case 1: {\em Compound symmetry (CS)}. $\Sigma(t,t^\prime) = \sigma^2 \{\rho + (1-\rho)\mathbb{I}(t=t^\prime)\}$, \\
\noindent Case 2: {\em Conditional Autoregressive of order 1, CAR(1)}. $\Sigma(t,t^\prime) = \sigma^2 0.5^{\lvert t - t^\prime\rvert}$ with $\sigma^2=1$, and \\
\noindent Case 3: {\em Non-stationary}. $\Sigma(t,t^\prime) $ is equal to the covariance of the process  $Z(t) := \xi_{i1}\sqrt{2}\sin(2 \pi t) +  \xi_{i2}\sqrt{2}\cos(2 \pi t)$ where $\xi_{i1} \overset{\text{iid}}{\sim} N(0,1)$ and $\xi_{i2} \overset{\text{iid}}{\sim} N(0,0.5)$. \\
\noindent We also investigate two different sparsity levels of the sampling design, {\em high} and {\em medium}. For stationary covariance Cases 1 and 2, we assign {\em high} sparsity to $m_i=5$, and {\em medium} sparsity to $m_i=8$ sampling points for each subject within the interval $[0,1]$. In the non-stationary covariance Case 3, the {\em high} sparsity level selects $m_i$ randomly from $\{4,5,6,7\}$, and {\em medium}  sparsity level selects $m_i$ from $\{8,9,\ldots,12\}$. The measurement error variance is set to $\sigma^2_e = 0.001$.

\begin{table}[ht]
\centering
\caption{Power function of the test for different sample sizes and magnitude of effect size.}
\scalebox{0.7}{
\begin{tabular}{clcccccc}
\multicolumn{8}{c}{Case 1 : {\em Compound symmetric}} \\ \hline \hline
\multicolumn{8}{c}{\# observations per subject: \textit{low}} \\ \hline
\multicolumn{2}{c}{\multirow{2}{*}{$n$}} & \multicolumn{2}{c}{$\eta = 0.5$} & \multicolumn{2}{c}{$\eta  = 0.75$} & \multicolumn{2}{c}{$\eta  = 1$} \\ \cline{3-8} 
\multicolumn{2}{c}{} & $\mathcal{P}_{n}(\eta)$ & \multicolumn{1}{l}{$\widehat{\mathcal{P}}_{n}(\eta)$} & $\mathcal{P}_{n}(\eta)$ & \multicolumn{1}{l}{$\widehat{\mathcal{P}}_{n}(\eta)$} & $\mathcal{P}_{n}(\eta)$ & \multicolumn{1}{l}{$\widehat{\mathcal{P}}_{n}(\eta)$} \\ \hline
\multicolumn{2}{c}{100} & 0.16 & \multicolumn{1}{c}{0.17} & 0.30 & \multicolumn{1}{c}{0.32} & 0.48 & 0.49 \\
\multicolumn{2}{c}{200} & 0.28 & \multicolumn{1}{c}{0.29} & 0.54 & \multicolumn{1}{c}{0.56} & 0.78 & 0.80 \\
\multicolumn{2}{c}{400} & 0.49 & \multicolumn{1}{c}{0.49} & 0.83 & \multicolumn{1}{c}{0.84} & 0.97 & 0.97 \\
\hline
\multicolumn{8}{c}{\# observations per subject: \textit{medium}} \\ \hline
\multicolumn{2}{c}{\multirow{2}{*}{$n$}} & \multicolumn{2}{c}{$\eta = 0.5$} & \multicolumn{2}{c}{$\eta  = 0.75$} & \multicolumn{2}{c}{$\eta  = 1$} \\ \cline{3-8} 
\multicolumn{2}{c}{} & $\mathcal{P}_{n}(\eta)$ & \multicolumn{1}{l}{$\widehat{\mathcal{P}}_{n}(\eta)$} & $\mathcal{P}_{n}(\eta)$ & \multicolumn{1}{l}{$\widehat{\mathcal{P}}_{n}(\eta)$} & $\mathcal{P}_{n}(\eta)$ & \multicolumn{1}{l}{$\widehat{\mathcal{P}}_{n}(\eta)$} \\ \hline
\multicolumn{2}{c}{100} & 0.16 & \multicolumn{1}{c}{0.17} & 0.29 & \multicolumn{1}{c}{0.33} & 0.47 & 0.51 \\
\multicolumn{2}{c}{200} & 0.27 & \multicolumn{1}{c}{0.30} & 0.51 & \multicolumn{1}{c}{0.56} & 0.76 & 0.81 \\
\multicolumn{2}{c}{400} & 0.47 & \multicolumn{1}{c}{0.52} & 0.81 & \multicolumn{1}{c}{0.85} & 0.96 & 0.98 \\
\hline 
\hline \\ 
\end{tabular}
} \quad
\scalebox{0.7}{
\begin{tabular}{clcccccc}
\multicolumn{8}{c}{Case 2 : {\em Conditional auto-regressive of order 1}} \\ \hline \hline
\multicolumn{8}{c}{\# observations per subject: \textit{low}} \\ \hline
\multicolumn{2}{c}{\multirow{2}{*}{$n$}} & \multicolumn{2}{c}{$\eta = 0.5$} & \multicolumn{2}{c}{$\eta  = 0.75$} & \multicolumn{2}{c}{$\eta  = 1$} \\ \cline{3-8} 
\multicolumn{2}{c}{} & $\mathcal{P}_{n}(\eta)$ & \multicolumn{1}{l}{$\widehat{\mathcal{P}}_{n}(\eta)$} & $\mathcal{P}_{n}(\eta)$ & \multicolumn{1}{l}{$\widehat{\mathcal{P}}_{n}(\eta)$} & $\mathcal{P}_{n}(\eta)$ & \multicolumn{1}{l}{$\widehat{\mathcal{P}}_{n}(\eta)$} \\ \hline
\multicolumn{2}{c}{100} & 0.47 & \multicolumn{1}{c}{0.32} & 0.82 & \multicolumn{1}{c}{0.59} & 0.98 & 0.75 \\
\multicolumn{2}{c}{200} & 0.79 & \multicolumn{1}{c}{0.59} & 0.98 & \multicolumn{1}{c}{0.82} & 1.00 & 0.89 \\
\multicolumn{2}{c}{400} & 0.98 & \multicolumn{1}{c}{0.85} & 1.00 & \multicolumn{1}{c}{0.93} & 1.00 & 0.95 \\
\hline
\multicolumn{8}{c}{\# observations per subject: \textit{medium}} \\ \hline
\multicolumn{2}{c}{\multirow{2}{*}{$n$}} & \multicolumn{2}{c}{$\eta = 0.5$} & \multicolumn{2}{c}{$\eta  = 0.75$} & \multicolumn{2}{c}{$\eta  = 1$} \\ \cline{3-8} 
\multicolumn{2}{c}{} & $\mathcal{P}_{n}(\eta)$ & \multicolumn{1}{l}{$\widehat{\mathcal{P}}_{n}(\eta)$} & $\mathcal{P}_{n}(\eta)$ & \multicolumn{1}{l}{$\widehat{\mathcal{P}}_{n}(\eta)$} & $\mathcal{P}_{n}(\eta)$ & \multicolumn{1}{l}{$\widehat{\mathcal{P}}_{n}(\eta)$} \\ \hline
\multicolumn{2}{c}{100} & 0.45 & \multicolumn{1}{c}{0.44} & 0.81 & \multicolumn{1}{c}{0.74} & 0.97 & 0.91 \\
\multicolumn{2}{c}{200} & 0.75 & \multicolumn{1}{c}{0.74} & 0.98 & \multicolumn{1}{c}{0.96} & 0.99 & 0.99 \\
\multicolumn{2}{c}{400} & 0.97 & \multicolumn{1}{c}{0.96} & 1.00 & \multicolumn{1}{c}{1.00} & 1.00 & 1.00 \\
\hline 
\hline \\
\end{tabular}
}
\scalebox{0.7}{
\begin{tabular}{clcccccc}
\multicolumn{8}{c}{Case 3 : {\em Non-stationary}} \\ \hline \hline
\multicolumn{8}{c}{\# observations per subject: \textit{low}} \\ \hline
\multicolumn{2}{c}{\multirow{2}{*}{$n$}} & \multicolumn{2}{c}{$\eta = 0.5$} & \multicolumn{2}{c}{$\eta  = 0.75$} & \multicolumn{2}{c}{$\eta  = 1$} \\ \cline{3-8} 
\multicolumn{2}{c}{} & $\mathcal{P}_{n}(\eta)$ & \multicolumn{1}{l}{$\widehat{\mathcal{P}}_{n}(\eta)$} & $\mathcal{P}_{n}(\eta)$ & \multicolumn{1}{l}{$\widehat{\mathcal{P}}_{n}(\eta)$} & $\mathcal{P}_{n}(\eta)$ & \multicolumn{1}{l}{$\widehat{\mathcal{P}}_{n}(\eta)$} \\ \hline
\multicolumn{2}{c}{100} & 0.08 & \multicolumn{1}{c}{0.08} & 0.12 & \multicolumn{1}{c}{0.13} & 0.17 & 0.19\\
\multicolumn{2}{c}{200} & 0.11 & \multicolumn{1}{c}{0.10} & 0.20 & \multicolumn{1}{c}{0.21} & 0.31 & 0.31 \\
\multicolumn{2}{c}{400} & 0.18 & \multicolumn{1}{c}{0.17} & 0.35 & \multicolumn{1}{c}{0.32} & 0.57 & 0.58 \\
\hline
\multicolumn{8}{c}{\# observations per subject: \textit{medium}} \\ \hline
\multicolumn{2}{c}{\multirow{2}{*}{$n$}} & \multicolumn{2}{c}{$\eta = 0.5$} & \multicolumn{2}{c}{$\eta  = 0.75$} & \multicolumn{2}{c}{$\eta  = 1$} \\ \cline{3-8} 
\multicolumn{2}{c}{} & $\mathcal{P}_{n}(\eta)$ & \multicolumn{1}{l}{$\widehat{\mathcal{P}}_{n}(\eta)$} & $\mathcal{P}_{n}(\eta)$ & \multicolumn{1}{l}{$\widehat{\mathcal{P}}_{n}(\eta)$} & $\mathcal{P}_{n}(\eta)$ & \multicolumn{1}{l}{$\widehat{\mathcal{P}}_{n}(\eta)$} \\ \hline
\multicolumn{2}{c}{100} & 0.08 & \multicolumn{1}{c}{0.08} & 0.12 & \multicolumn{1}{c}{0.12} & 0.17 & 0.20 \\
\multicolumn{2}{c}{200} & 0.11 & \multicolumn{1}{c}{0.11} & 0.20 & \multicolumn{1}{c}{0.22} & 0.33 & 0.34 \\
\multicolumn{2}{c}{400} & 0.18 & \multicolumn{1}{c}{0.17} & 0.37 & \multicolumn{1}{c}{0.37} & 0.58 & 0.58 \\
\hline 
\hline
\end{tabular}
}
\label{tab: PW_CS}
\end{table}

\subsection{Validation of power function} \label{sec: simpowerfunction}

We employ Algorithm~\ref{alg: powerfunction} to calculate the test's power function for various total sample sizes, $n$, and effect sizes, $\eta$ specified in the mean difference function, under the Cases specified in Section~\ref{sec: edfd_testing_simstudy:dataGen}. In these simulation studies, we assume equal sample allocations in both groups ($\kappa=1$). For notational ease, we suppress the subscript $\kappa$ in the theoretical power function of the test and denote it by $\mathcal{P}_n(\eta)$. Table~\ref{tab: PW_CS} displays the theoretical power for $n=100, 200$, and $400$, and $\eta = 0.5, 0.75$, and $1$, considering different sparsity levels and covariance structures. The results indicate that the power function increases as the sample size and group differences grow. Additionally, we observe variations in the power function across different covariance structures, while the sparsity level seems to have minimal influence. This suggests that precise estimation of eigenfunctions and population covariance of scores mitigates the impact of sparsity on the power function. However, in cases with small sample sizes, denser designs yield better eigenfunction estimation, leading to more accurate power function calculations.

To validate the theoretical power, we conduct Monte-Carlo simulations. For fixed $n$, specified mean difference $\eta$, covariance structure, and random sampling design, we generate $1000$ replications. We then calculate the percentage of times the test rejects the null hypothesis using test rule~(\ref{eqn: HotellingTestRule_eqvar}) to obtain the empirical power denoted as $\widehat{\mathcal{P}}_n(\eta)$. This empirical power is presented alongside the theoretically calculated power function ${\mathcal{P}}_n(\eta)$ in Table~\ref{tab: PW_CS}. The results show a close match between the power function and the empirically calculated power, especially when a larger number of observations per subject is available. For instance, when using an AR(1) covariance structure with only five observations per subject, the eigenfunctions are not well estimated, leading to a lower empirical power. However, with eight observations per subject, the two power measures align more closely. This highlights the impact of low sparsity on empirical power and underscores the importance of precise covariance component estimation based on a large number of subjects (Step 1 of Algorithm~\ref{alg: powerfunction}) for reliable power function estimation.

\subsection{Validation of missing-immunity of the test} \label{sec: sim_missing_immune}
In the previous subsection, we numerically validated the efficiency of Algorithm~\ref{alg: powerfunction} for calculating the theoretical power function. Now, we present the power of the test for various values of $n$ and different percentages of missing observations per subject in Table~\ref{tab: Miss_CS} to assess the test's robustness to increasing missing data.

As shown, the power of the test remains relatively stable even as the percentage of missing observations per subject increases from $0\%$ to as high as $40\%$. This phenomenon underscores the test's ``missing-immunity". This is attributed to consistent estimation of eigencomponents based on a large sample size, as documented in Step 1 of Algorithm~\ref{alg: sampsize}. However, it  is worth noting that if the missing percentage becomes so high that the number of observations per subject effectively drops to extremely low levels (less than three), the subject-specific `shrinkage' scores and their covariances $\{\widehat{\boldsymbol\Lambda}_g\}_{g=1}^2$ cannot be reliably estimated, thereby affecting the test's power function.

\begin{table}[ht]
\centering
\caption{Power function of test across different sample size and missing percentages.}
\label{tab: Miss_CS}
\scalebox{0.75}{
\begin{tabular}{ccccccccc}
\multicolumn{9}{c}{Case 1 : {\em Compound symmetric}} \\
\hline
\hline
\multicolumn{9}{c}{\# observations per subject : {\em low}}                                                          \\ \hline
\multicolumn{1}{c}{\multirow{2}{*}{\begin{tabular}[c]{@{}c@{}}$n$\end{tabular}}} & \multicolumn{4}{c}{$\eta = 0.5$} & \multicolumn{4}{c}{$\eta = 0.75$} \\ \cline{2-9} 
\multicolumn{1}{c}{}           & $0\%$ & $10\%$ & $20\%$ & \multicolumn{1}{c}{$40\%$} & $0\%$ & $10\%$ & $20\%$ & $40\%$ \\ \hline
%\multicolumn{1}{c}{$200$}   & 0.15 & 0.10   & 0.10   & \multicolumn{1}{c}{0.13}  & 0.27 & 0.26  & 0.33  & 0.34  \\
\multicolumn{1}{c}{$ 100$}  & 0.16 & 0.16   & 0.16   & \multicolumn{1}{c}{0.17}  & 0.30 & 0.30  & 0.31  & 0.32  \\
\multicolumn{1}{c}{$ 200$}  & 0.28 & 0.28   & 0.28   & \multicolumn{1}{c}{0.30}  & 0.54 & 0.54  & 0.54  & 0.57  \\
\multicolumn{1}{c}{$ 400$} & 0.49 & 0.49   & 0.50   & \multicolumn{1}{c}{0.53}  & 0.83 & 0.83  & 0.83  & 0.86 \\ \hline

\multicolumn{9}{c}{\# observations per subject : {\em medium}}                                                          \\ \hline
\multicolumn{1}{c}{\multirow{2}{*}{\begin{tabular}[c]{@{}c@{}}$n$\end{tabular}}} & \multicolumn{4}{c}{$\eta = 0.5$} & \multicolumn{4}{c}{$\eta = 0.75$} \\ \cline{2-9} 
\multicolumn{1}{c}{}           & $0\%$ & $10\%$ & $20\%$ & \multicolumn{1}{c}{$40\%$} & $0\%$ & $10\%$ & $20\%$ & $40\%$ \\ \hline
%\multicolumn{1}{c}{$200$}   & 0.17 & 0.14   & 0.10   & \multicolumn{1}{c}{0.14}  & 0.57 & 0.41  & 0.54  & 0.54  \\
\multicolumn{1}{c}{$ 100$}  & 0.16 & 0.15   & 0.16  & \multicolumn{1}{c}{0.16}  & 0.29 & 0.29  & 0.29  & 0.30  \\
\multicolumn{1}{c}{$ 200$}  & 0.27 & 0.26   & 0.27   & \multicolumn{1}{c}{0.28}  & 0.51 & 0.51  & 0.51  & 0.54  \\
\multicolumn{1}{c}{$ 400$} & 0.47 & 0.47   & 0.47   & \multicolumn{1}{c}{0.50}  & 0.81 & 0.80  & 0.81  & 0.83 \\ \hline 
\end{tabular}
} \qquad
\scalebox{0.75}{
\begin{tabular}{ccccccccc}
\multicolumn{9}{c}{Case 2 : {\em Conditional auto-regressive of order $1$}} \\
\hline
\hline
\multicolumn{9}{c}{\# observations per subject : {\em low}}                                                          \\ \hline
\multicolumn{1}{c}{\multirow{2}{*}{\begin{tabular}[c]{@{}c@{}}$n$\end{tabular}}} & \multicolumn{4}{c}{$\eta = 0.5$} & \multicolumn{4}{c}{$\eta = 0.75$} \\ \cline{2-9} 
\multicolumn{1}{c}{}           & $0\%$ & $10\%$ & $20\%$ & \multicolumn{1}{c}{$40\%$} & $0\%$ & $10\%$ & $20\%$ & $40\%$ \\ \hline
%\multicolumn{1}{c}{$200$}   & 0.08 & 0.07   & 0.07   & \multicolumn{1}{c}{0.07}  & 0.12 & 0.12  & 0.12  & 0.11  \\
\multicolumn{1}{c}{$ 100$}  & 0.47 & 0.48  & 0.49   & \multicolumn{1}{c}{0.56}  & 0.82 & 0.83  & 0.85  & 0.87  \\
\multicolumn{1}{c}{$ 200$}  & 0.79 & 0.80   & 0.82   & \multicolumn{1}{c}{0.87}  & 0.98 & 0.99   & 0.99  & 1.00  \\
\multicolumn{1}{c}{$ 400$} & 0.98 & 0.98   & 0.99  & \multicolumn{1}{c}{1.00}  & 1.00 & 1.00  & 1.00  & 1.00 \\ \hline

\multicolumn{9}{c}{\# observations per subject : {\em medium}}                                                          \\ \hline
\multicolumn{1}{c}{\multirow{2}{*}{\begin{tabular}[c]{@{}c@{}}$n$\end{tabular}}} & \multicolumn{4}{c}{$\eta = 0.5$} & \multicolumn{4}{c}{$\eta = 0.75$} \\ \cline{2-9} 
\multicolumn{1}{c}{}           & $0\%$ & $10\%$ & $20\%$ & \multicolumn{1}{c}{$40\%$} & $0\%$ & $10\%$ & $20\%$ & $40\%$ \\ \hline
%\multicolumn{1}{c}{$200$}   & 0.15 & 0.15   & 0.07   & \multicolumn{1}{c}{0.07}  & 0.53 & 0.37  & 0.21  & 0.12 \\
\multicolumn{1}{c}{$ 100$}  & 0.45 & 0.45  & 0.45   & \multicolumn{1}{c}{0.49}  & 0.81 & 0.81   & 0.81  & 0.84  \\
\multicolumn{1}{c}{$ 200$}  & 0.75 & 0.76   & 0.76   & \multicolumn{1}{c}{0.81}  & 0.98 & 0.98   & 0.98  & 0.99  \\
\multicolumn{1}{c}{$ 400$} & 0.97 & 0.97   & 0.98  & \multicolumn{1}{c}{0.98}  & 1.00 & 1.00  & 1.00  & 1.00 \\ \hline
\end{tabular}
}
\end{table}

\subsection{Sample size validation}
Table~\ref{tab: SS_CS} displays the minimum sample size required to attain a $100\gamma\%$ power for different values of the group difference parameter $\eta$, ranging from $0.5$ to $2$. This calculation is performed across all three covariance structures, assuming equal allocation of samples in each group $(\kappa=1)$. The combined sample size for both groups is determined using Algorithm~\ref{alg: sampsize} and is presented in the column labeled $n_{\min}$ in the tables. To validate these calculated minimum sample sizes, we conduct 1000 simulated experiments with the specified sample size, incorporating the designated mean difference ($\eta$) and underlying covariance structure. In each simulation, we apply our projection-based test and compute the empirical power, $\widehat{P}_{n_{min}}(\eta)$, which is reported in the second column of the corresponding cell. As the sample size increases, we observe a growing alignment between empirical power and the desired target power $\gamma$, affirming the accuracy of our test's sample size calculation algorithm.

\begin{table}
\centering
\caption{Minimum sample size required to achieve a target power of $100\gamma\%$.}
\scalebox{0.7}{
\begin{tabular}{clcccccc}
\multicolumn{8}{c}{Case 1 : {\em Compound symmetric}} \\ \hline 
\hline
\multicolumn{8}{c}{\# observations per subject: \textit{medium}} \\ \hline
\multicolumn{2}{c}{\multirow{2}{*}{$\eta$}} & \multicolumn{2}{c}{$\gamma  = 0.7$} & \multicolumn{2}{c}{$\gamma  = 0.8$} & \multicolumn{2}{c}{$\gamma  = 0.9$} \\ \cline{3-8} 
\multicolumn{2}{c}{} & $n_{\min}$ & \multicolumn{1}{l}{$\widehat{\mathcal{P}}_{n_{\min}}(\eta)$} & $n_{\min}$ & \multicolumn{1}{l}{$\widehat{\mathcal{P}}_{n_{\min}}(\eta)$} & $n_{\min}$ & \multicolumn{1}{l}{$\widehat{\mathcal{P}}_{n_{\min}}(\eta)$} \\ \hline
%\multicolumn{2}{c}{0.5} & 690 & \multicolumn{1}{c}{0.67} & 876 & \multicolumn{1}{c}{0.77} & 1179 & 0.86 \\
\multicolumn{2}{c}{0.75} & 296 & \multicolumn{1}{c}{0.75} & 390 & \multicolumn{1}{c}{0.83} & 528 & 0.93 \\
\multicolumn{2}{c}{1} & 172 & \multicolumn{1}{c}{0.74} & 226 & \multicolumn{1}{c}{0.82} & 296 & 0.93 \\
\hline
\hline \\
\end{tabular}
} \quad 
\scalebox{0.7}{
\begin{tabular}{clcccccc}
\multicolumn{8}{c}{Case 2 : {\em Conditional auto-regressive of order 1}} \\ \hline
 \hline
\multicolumn{8}{c}{\# observations per subject: \textit{medium}} \\ \hline
\multicolumn{2}{c}{\multirow{2}{*}{$\eta$}} & \multicolumn{2}{c}{$\gamma  = 0.7$} & \multicolumn{2}{c}{$\gamma  = 0.8$} & \multicolumn{2}{c}{$\gamma  = 0.9$} \\ \cline{3-8} 
\multicolumn{2}{c}{} & $n_{\min}$ & \multicolumn{1}{l}{$\widehat{\mathcal{P}}_{n_{\min}}(\eta)$} & $n_{\min}$ & \multicolumn{1}{l}{$\widehat{\mathcal{P}}_{n_{\min}}(\eta)$} & $n_{\min}$ & \multicolumn{1}{l}{$\widehat{\mathcal{P}}_{n_{\min}}(\eta)$} \\ \hline
%\multicolumn{2}{c}{0.5} & 178 & \multicolumn{1}{c}{0.66} & 222 & \multicolumn{1}{c}{0.78} & 292 & 0.89 \\
\multicolumn{2}{c}{0.75} & 82 & \multicolumn{1}{c}{0.65} & 100 & \multicolumn{1}{c}{0.74} & 130 & 0.87 \\
\multicolumn{2}{c}{1} & 54 & \multicolumn{1}{c}{0.68} & 60 & \multicolumn{1}{c}{0.75} & 82 & 0.85 \\
\hline
\hline  \\
\end{tabular}
} 
\scalebox{0.7}{
\begin{tabular}{clcccccc}
\multicolumn{8}{c}{Case 3 : {\em Non-stationary}} \\ \hline
\hline
\multicolumn{8}{c}{\# observations per subject: \textit{medium}} \\ \hline
\multicolumn{2}{c}{\multirow{2}{*}{$\eta$}} & \multicolumn{2}{c}{$\gamma  = 0.7$} & \multicolumn{2}{c}{$\gamma  = 0.8$} & \multicolumn{2}{c}{$\gamma  = 0.9$} \\ \cline{3-8} 
\multicolumn{2}{c}{} & $n_{\min}$ & \multicolumn{1}{l}{$\widehat{\mathcal{P}}_{n_{\min}}(\eta)$} & $n_{\min}$ & \multicolumn{1}{l}{$\widehat{\mathcal{P}}_{n_{\min}}(\eta)$} & $n_{\min}$ & \multicolumn{1}{l}{$\widehat{\mathcal{P}}_{n_{\min}}(\eta)$} \\ \hline
%\multicolumn{2}{c}{0.5} & 2002 & \multicolumn{1}{c}{0.70} & 2509 & \multicolumn{1}{c}{0.80} & 3274 & 0.90 \\
\multicolumn{2}{c}{0.75} & 868 & \multicolumn{1}{c}{0.69} & 1060 & \multicolumn{1}{c}{0.82} & 1430 & 0.89 \\
\multicolumn{2}{c}{1} & 496 & \multicolumn{1}{c}{0.70} & 618 & \multicolumn{1}{c}{0.80} & 800 & 0.89 \\
\hline
\hline
\end{tabular}
}
\label{tab: SS_CS}
\end{table}

\section{Application on real data} \label{sec: realdata}

\subsection{Azillect study}
In the Azillect clinical trial, a Phase 3 study conducted in Japan, the primary endpoints for early Parkinson's disease (PD) patients were assessed using the Movement Disorder Society Unified Parkinson Disease Rating Scale (MDS-UPDRS) Parts II and III scores, where a higher score indicates greater disease severity. The trial randomized participants aged 30-79 years, diagnosed with PD within the past five years, into two groups: one receiving the study drug rasagiline and the other receiving a placebo for up to 26 weeks. 244 subjects were recruited for this study, with $118$ subjects in the rasagiline group and 126 subjects in the placebo group. The participants were assessed
at baseline, and at week 6, 10, 14, 20 and 26 thereafter. More information can be found in \cite{hattori2019rasagiline}. A comparison of key PD characteristics between the two groups are presented in Table~\ref{tab:azillect_baseline} of the Supplement. We will demonstrate how the projection-based test can be applied to Azillect study and how findings from the analysis can inform  design of a new study.

\begin{figure}
 \centering
\includegraphics[scale=0.47]{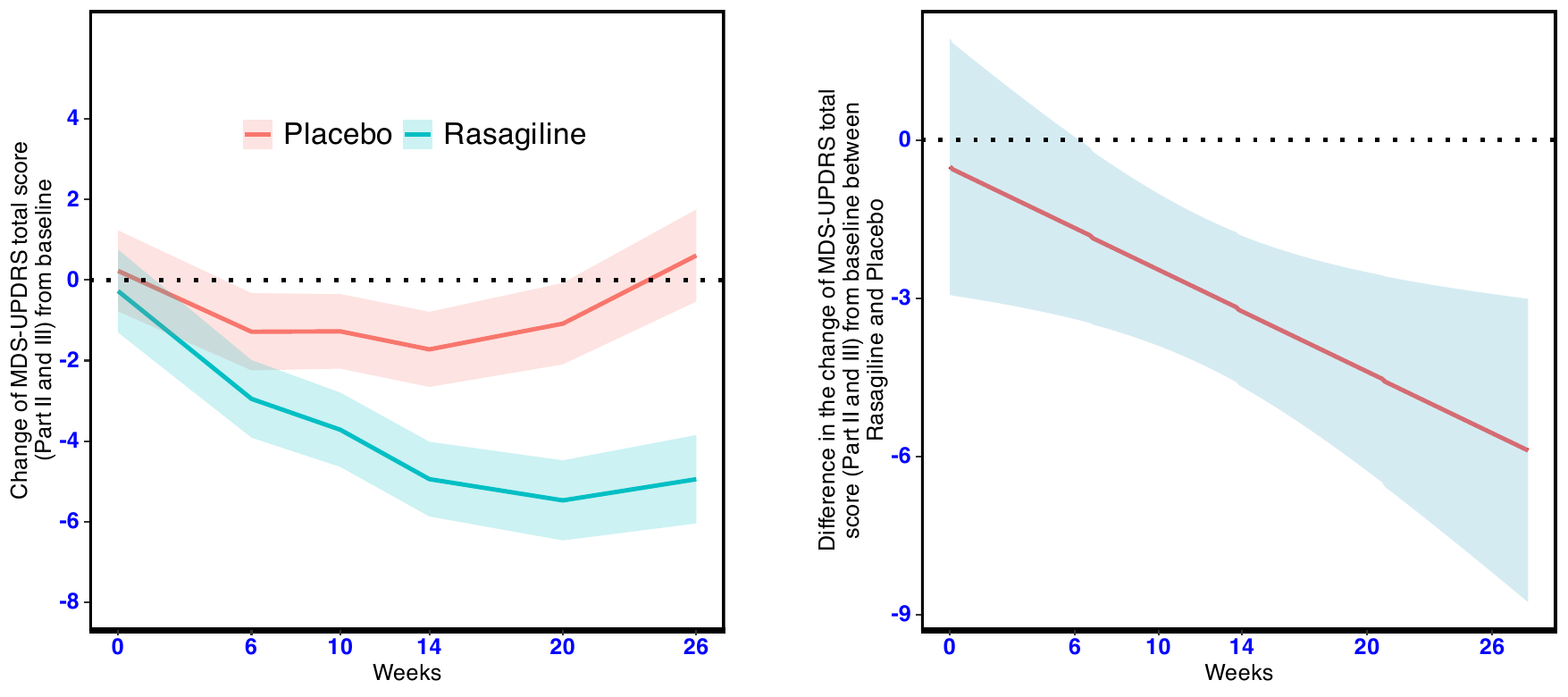}
\cprotect\caption{Findings from Azillect study.  \textsl{Left panel}: average change of MDS-UPDRS Parts II-III total 
score from baseline for rasagiline and placebo group (indicated by solid lines). The shaded region reflects the $95\%$ pointwise confidence interval. \textsl{Right panel}: Estimated difference between the average change of MDS-UPDRS Parts II-III total score from baseline between the two groups, along with $95\%$ pointwise confidence interval (shaded in blue).}
\label{fig: effect_azillect}
\end{figure}

At the end of the 26-week study, the mean change in the MDS-UPDRS Parts II-III total score from baseline was $1.87$ for the placebo group and $-4.52$ for the rasagiline group. The left panel of Figure~\ref{fig: effect_azillect} illustrates the average change in the total Parts II-III scores from baseline for both the rasagiline and placebo groups. This illustration is based on fitting model~(\ref{eqn: dgm}) using the \verb|gam()| function in the \verb|mgcv| \citep{wood2011fast} package in R. The right panel of Figure~\ref{fig: effect_azillect} displays the average change in MDS-UPDRS Parts II and III total scores from baseline, specifically comparing the rasagiline and placebo groups. The solid red line represents the estimate of the treatment effect, which is a smooth estimate of $\widehat{\mu}_{\text{rasagiline}}(t) - \widehat{\mu}_{\text{placebo}}(t)$. It is accompanied by a $95\%$ pointwise confidence interval shaded in blue. The descending trajectory of the solid red line illustrates rasagiline's effectiveness in slowing PD's progression compared to placebo.

The projection-based test allows for non-parametric modeling of the data's covariance structure. The left panel of Figure~\ref{fig: efunctions} (in Section~\ref{sec:appendixefunctions} of Supplement) displays the two leading eigenfunctions obtained from functional Principal Component Analysis (fPCA) on the estimated covariance. The eigenvalues corresponding to the two eigenfunctions are $31.8$ (represented by the red line) and $2.6$ (represented by the blue line), indicating that the first eigenfunction explains $91\%$ of the total data variability. The pattern of these eigenfunctions provides valuable insights into the nature of score variations. The first eigenfunction (in red) exhibits a relatively stable profile over time, suggesting that the largest variation among subjects relates to their subject-specific average changes in MDS-UPDRS Parts II and III total scores from the baseline. This implies that a random subject-specific intercept effectively captures the major source of variation in the data.

Using the estimated eigenfunctions, we compute the subject-specific fPC scores for both groups to conduct the Hotelling-T$^2$ test. The confidence band (not covering $0$) in the right panel of Figure~\ref{fig: effect_azillect}, along with a low p-value ($< 3 \times 10^{-6}$) from the projection-based test, provides strong evidence that rasagiline significantly reduces the MDS-UPDRS Parts II-III total scores over the $26$ weeks compared to the placebo. If we intend to design a new study aiming for the treatment effect observed in the right panel of Figure~\ref{fig: effect_azillect}, along with the covariance structure specified by the two leading eigenfunctions from the left panel of Figure~\ref{fig: efunctions}, we would require a minimum sample size of $74$ for an $80\%$ power level or $92$ for a $90\%$ power level, assuming an equal allocation ratio between the two groups. This sample size is comparable to what would be needed for a t-test, which is around 90 for a target power of $90\%$, based on the effect size observed at the end of week 26. The reason for this similarity is that the t-test relies on the estimated effect size at the end of the trial when the effect size curve reaches its maximum (as shown in the right panel of Figure~\ref{fig: effect_azillect}). However, in a scenario where the treatment effect peaks during the middle of the study and diminishes towards the trial's conclusion, the projection-based test would exhibit significantly higher power compared to the t-test, which only considers the effect size at the trial's end.

\subsection{SURE-PD3 study}
In the SURE-PD3 trial, a randomized, double-blind, placebo-controlled study, researchers aimed to evaluate whether inosine could slow down the progression of Parkinson's Disease in its Phase 3 stage. The trial focused on individuals in the early stages of Parkinson's Disease and investigated whether urate-elevating inosine treatment could reduce the clinical decline. The primary measure was the rate of change in the MDS-UPDRS Parts I, II, and III total score, a scale combining patient and clinician-reported outcomes. Higher scores indicated more severe disability, with a range from 0 to 236 points (See \cite{bluett2021effect} for detailed information on the trial). In this study, a total of 298 participants were enrolled, equal allocation in each of the Inosine and Placebo group. Assessments were conducted initially at baseline and at week 3, week 6, month 3, and  subsequently at every 3 months for 24 months. For a detailed comparison of essential Parkinson's disease (PD) characteristics between the two groups, refer to Table~\ref{tab:baseline_surepd3} in the Supplement. We will apply the projection-based test on this study and demonstrate how to design a new clinical trial using our PASS toolbox based on the findings from the SURE-PD3 trial.

\begin{figure}
 \centering
 \includegraphics[scale=0.47]{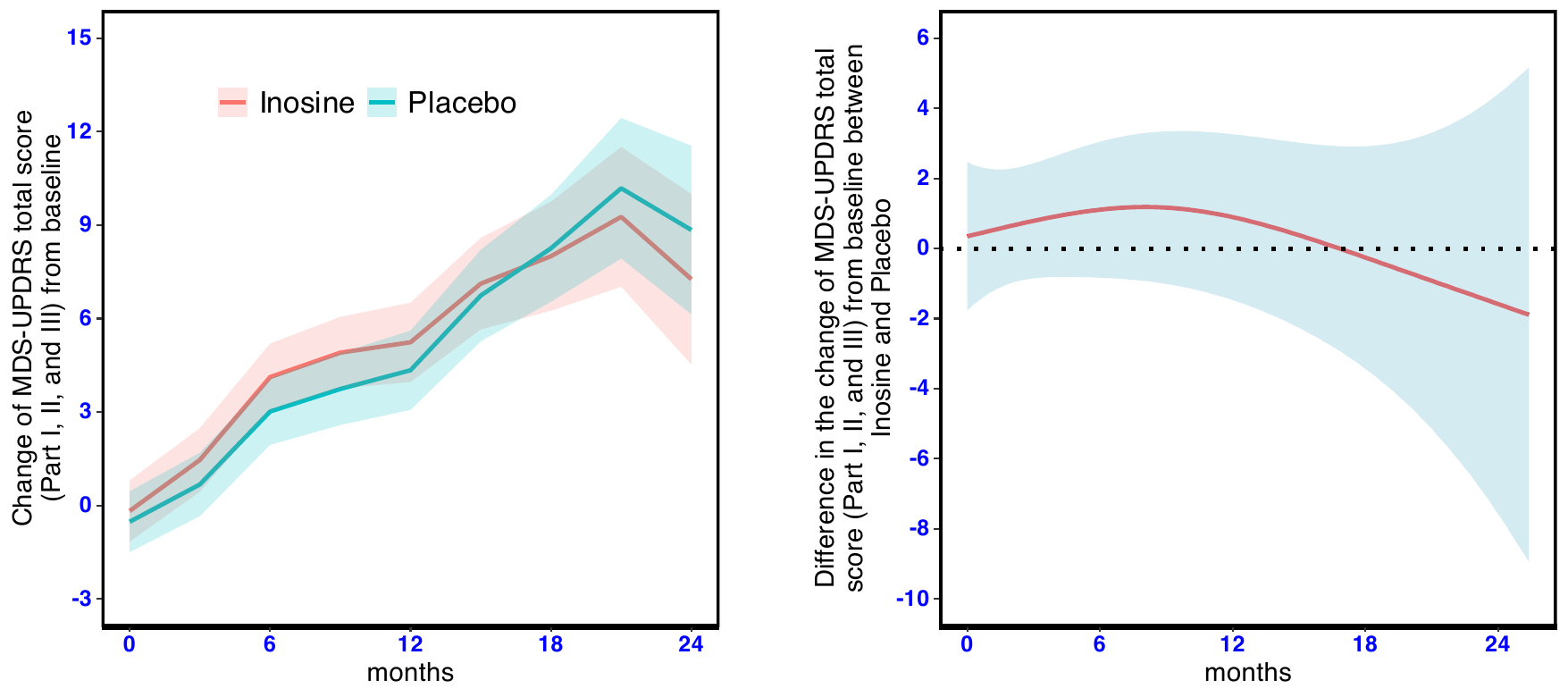}
 \cprotect\caption{Findings from SURE-PD3 study.  \textsl{Left panel}: average change of MDS-UPDRS Parts I-III total 
score from baseline for inosine and placebo group (indicated by solid lines). The shaded region reflects the $95\%$ pointwise confidence interval. \textsl{Right panel}: Estimated difference between the average change of MDS-UPDRS Parts I-III total score from baseline between the two groups, along with $95\%$ pointwise confidence interval (shaded in blue).}
 \label{fig: effect_surepd3}
 \end{figure}

The left panel of Figure~\ref{fig: effect_surepd3} displays the estimated change in total MDS-UPDRS Parts I-III scores from baseline for both the inosine group (in red) and the placebo group (in blue), along with the $95\%$ pointwise confidence band. The analysis does not reveal a significant difference in the change of total scores from baseline between the two groups. The right panel of Figure~\ref{fig: efunctions} (in Section~\ref{sec:appendixefunctions} of Supplement) presents the two leading eigenfunctions obtained from conducting fPCA on the smoothed estimated covariance. The first eigenfunction explains $91\%$ of the total variability. Similar to the Azillect study, the largest variation (in red) among subjects is captured by a random subject-specific intercept term.

After extracting the eigencomponents through fPCA, we use subject-specific fPC scores from both groups to conduct the Hotelling-T$^2$ test. The p-value of the Hotelling T$^2$ test is $0.42$, indicating that inosine does not significantly improve the total MDS-UPDRS scores compared to the placebo over 24 months. This conclusion is further supported by the fact that the $95\%$ pointwise confidence interval of the mean difference, obtained from the \verb|gam()| (shown in the right panel of Figure~\ref{fig: effect_surepd3}), contains zero. Additionally, the length of the confidence band increases in the later months due to a considerable number of early withdrawals.

If we design a new study where the effect of study drug follows the pattern of the red line as in the right panel of Figure~\ref{fig: effect_surepd3}, indicating a positive treatment effect after approximately 18 months, and the covariance of the data is characterized by the two principal eigenfunctions (as shown in the right panel of Figure~\ref{fig: efunctions}), we would need approximately $232$ subjects in each group to achieve a power of $80\%$ power level and about $286$ subjects in each group for a power of $90\%$ power level.

\section{Conclusion and discussion}

In conclusion, this article introduces a novel PASS formula for the projection-based Hotelling $T^2$ type test, designed specifically for sparse and irregularly observed functional data. Building upon the work by \cite{wang2021two}, the  distribution of the test-statistic closely approximates a non-central $\chi^2$ distribution under the alternate hypothesis. This article develops the test's power function and offers an efficient algorithm, available in the R package \verb|fPASS| \citep{fPASS2023}, to determine the minimum sample size required to achieve a target power by inversely mapping the power function.

The significance of the PASS formula lies in its adaptability. It accommodates various group difference structures without the need for predefined specific parametric forms, offers flexibility in specifying non-stationary smooth covariance structures for responses, and handles varying observation times and counts per subject. Numerical studies demonstrate the test's robustness in power even with increasing missing observations within a certain range, making it suitable for real-world applications where missing data is common.

Furthermore, the projection-based test can be extended to compare means across more than two groups, similar to a multivariate analysis of variance (MANOVA). Various MANOVA test statistics, such as Wilk's Lambda \citep{wilks1932certain}, Pillai's trace \citep{pillai1955some}, can be derived using the latent roots of the matrix constructed by multiplying the between-group sum of squares (SSQ) matrix and the inverse of within-group SSQ matrix in the MANOVA table. However, extending the PASS formula for this scenario requires deriving the distribution of MANOVA test statistics under the alternate hypothesis. This a complex task because the covariances of the `shrinkage' scores under the alternate hypothesis differ across groups. A starting point for such research could be the work by \cite{muirhead2009aspects}, which outlines the asymptotic distribution of MANOVA statistics under equal variances. This research direction holds promise for future investigations, especially concerning clinical trials involving more than two treatment arms.

\section*{Funding statement}
Support for this work comes from National Institute on Aging (grants AG064803 and P30AG072958) to Sheng Luo.

\spacingset{0.95}
\bibliographystyle{agsm}
\bibliography{refs_main}

\newpage

\renewcommand{\thesection}{S\arabic{section}}
\renewcommand{\thefigure}{F\arabic{figure}}
\renewcommand{\thetable}{T\arabic{table}}

\begin{center}

{\bf \Large Supplementary material for ``Power and sample size calculation for two-sample projection-based testing of sparsely observed functional data"}
\end{center}

\spacingset{1.45}

This supplementary material is divided into two parts. In section \ref{sec: proof_of_theorems}, we present the proof of Theorem~\ref{thm: power_egnknown}, and Proposition~\ref{thm: power_egnunknown}.  Additional results demonstrating the pattern of eigenfunctions obtained from the estimated covariance function of Azillect and SURE-PD3 study is presented Section~\ref{sec:appendixefunctions}.

\section{Proof of theorems} \label{sec: proof_of_theorems}

\subsection{Proof of Theorem~\ref{thm: power_egnknown}}

We first provide the proof of Theorem~\ref{thm: power_egnknown}.  Note that the proof is immediate if we can derive the alternate distribution of the Hotelling $T^2$ random variable defined in~(\ref{eqn: HotellingT2}) under the unequal variance assumption of the `shrinkage' scores.  The test statistic is of the form
$$ T_{n}  = \frac{n_1n_2}{n_1+n_2} (\widetilde{\boldsymbol\zeta}_{1+} - \widetilde{\boldsymbol\zeta}_{2+})^\top \widetilde{\boldsymbol\Lambda}^{-1} (\widetilde{\boldsymbol\zeta}_{1+} - \widetilde{\boldsymbol\zeta}_{2+}).$$ It is crucial to notice that when the true mean difference $\eta(t) = \mu_1(t) -\mu_2(t)$ is different from zero. From the formula of the mean and the covariance of the {\em shrinkage} scores (Eqn. (2.11) of \cite{wang2021two}), it is clear that the population covariance of the {\em shrinkage} scores are different when the true mean function of the two groups are different. On the other hand, under the null, the covariance matrices of the scores of the two groups are the same. Therefore, under the alternative we have to find the distribution of the Hotelling $T^2$ statistic under the assumption that the covariances of the scores between the two groups are different. Therefore, the alternate distribution of the test-statistic no more follows a non-central $F$ distribution. Here, we will establish the alternate distribution of $T_n$ assuming that the covariances of the scores between the two groups, $\Lambda_1$ and $\Lambda_2$ are different.  Under the assumption of normality of the {\em shrinkage} scores, 
\begin{equation*}
\widetilde{\boldsymbol\zeta}_{1+} - \widetilde{\boldsymbol\zeta}_{2+} \sim \text{N}_K(\boldsymbol\Delta, \boldsymbol\Lambda_1/n_1 + \boldsymbol\Lambda_2/n_2),
\end{equation*}
and
$$
\begin{aligned}
(n_g - 1)\widetilde{\boldsymbol\Lambda}_g \sim \textbf{W}_K(n_g - 1,\boldsymbol\Lambda_g) \qquad g=1,2. 
%\\(n_2 - 1)\widetilde{\boldsymbol\Lambda}_2 \sim \textbf{W}_K(n_2 - 1,\boldsymbol\Lambda_2)
\end{aligned}
$$
To scale things properly as a function of the sample size, we want to represent the distribution of $T_n$ in terms of the allocation ratio of the sample size between the two groups. Let $\kappa = n_1/n_2$ be the allocation ratio, then define, $\boldsymbol\Lambda^{\dagger} = \boldsymbol\Lambda_1 + \kappa\boldsymbol\Lambda_2$, so that $\widetilde{\boldsymbol\zeta}_{1+} - \widetilde{\boldsymbol\zeta}_{2+} \sim \text{N}_K(\boldsymbol\Delta, n_1^{-1}\boldsymbol\Lambda^{\dagger})$. We can represent $(1/n_1 + 1/n_2) T_n$ equivalently as 
\begin{align}
\nonumber (1/n_1 + 1/n_2) T_n &= n_1 (\widetilde{\boldsymbol\zeta}_{1+} - \widetilde{\boldsymbol\zeta}_{2+})^\top {\boldsymbol\Lambda^{\dagger}}^{-1/2} \\
& (n_1{\boldsymbol\Lambda^{\dagger}}^{-1/2}\widetilde{\boldsymbol\Lambda}{\boldsymbol\Lambda^{\dagger}}^{-1/2})^{-1} 
 {\boldsymbol\Lambda^{\dagger}}^{-1/2} (\widetilde{\boldsymbol\zeta}_{1+} - \widetilde{\boldsymbol\zeta}_{2+}) \label{eqn: supp_meat_rep}.
\end{align} 
Define $\mathcal{Z} = \sqrt{n_1} {\boldsymbol\Lambda^{\dagger}}^{-1/2} (\widetilde{\boldsymbol\zeta}_{1+} - \widetilde{\boldsymbol\zeta}_{2+})$ and  $ \mathcal{S} := n_1{\boldsymbol\Lambda^{\dagger}}^{-1/2}\widetilde{\boldsymbol\Lambda}{\boldsymbol\Lambda^{\dagger}}^{-1/2}$, so that $(1/n_1 + 1/n_2) T_n =   \mathcal{Z}^\top \mathcal{S}^{-1}\mathcal{Z}$.   Now, we derive the asymptotic distribution of $\mathcal{S}$. Note that,
\begin{align*}
\textstyle (n_1 + n_2 - 2)\mathcal{S} &= n_1{\boldsymbol\Lambda^{\dagger}}^{-1/2}\{(n_1-1)\widetilde{\boldsymbol\Lambda}_1 \}{\boldsymbol\Lambda^{\dagger}}^{-1/2} \\
 & \quad + n_1{\boldsymbol\Lambda^{\dagger}}^{-1/2}\{(n_2-1)\widetilde{\boldsymbol\Lambda}_2\}{\boldsymbol\Lambda^{\dagger}}^{-1/2}.
\end{align*}
By the property of Wishart distribution,
\begin{align*}
&n_1{\boldsymbol\Lambda^{\dagger}}^{-1/2}\{(n_1 - 1)\widetilde{\boldsymbol\Lambda}_1\}{\boldsymbol\Lambda^{\dagger}}^{-1/2} \sim \textbf{W}_K(n_1 - 1, n_1 \boldsymbol\Omega) \\
&n_1{\boldsymbol\Lambda^{\dagger}}^{-1/2}\{(n_2 - 1)\widetilde{\boldsymbol\Lambda}_2\}{\boldsymbol\Lambda^{\dagger}}^{-1/2} \sim \textbf{W}_K(n_2 - 1, n_2 (\mathbf{I}_K - \boldsymbol\Omega) ),
\end{align*}
where $\boldsymbol\Omega = {\boldsymbol\Lambda^{\dagger}}^{-1/2}\boldsymbol\Lambda_1{\boldsymbol\Lambda^{\dagger}}^{-1/2}$, and using the fact that $\boldsymbol\Lambda_2 = \kappa^{-1}(\boldsymbol\Lambda^{\dagger} - \boldsymbol\Lambda_1)$. By the result of sum of two Wishart distributions proved by \cite{nel1986solution}, $\mathcal{S}$ has an approximate Wishart distribution as follows,
\begin{align}
(n_1 + n_2 - 2)\mathcal{S}  \;\overset{\text{a. d}}{=}\;\; \mathbf{W}_K(\nu, \nu^{-1}\boldsymbol\Omega^*)  \label{eqn: supp_wishart_part},
\end{align} 
where \begin{align*}
 \boldsymbol\Omega^* &= n_1(n_1 -1) \boldsymbol\Omega + n_2 (n_2 -1)(\mathbf{I}_K -  \boldsymbol\Omega). \\
 &= n_2^2 \left\{\kappa (\kappa - 1/n_2)  \boldsymbol\Omega + (1-1/n_2)(\mathbf{I}_K -  \boldsymbol\Omega)\right\} \\
& = n_2^2 \boldsymbol\Omega^\dagger,
\end{align*} and  $\nu$ is the approximate degree of freedom of the Wishart distribution specified by  
\begin{align*}
\nu &= n_2 \left\{\tr({\boldsymbol\Omega^\dagger}^2) + \tr^2(\boldsymbol\Omega^\dagger)\right\}  \left[\kappa^2(\kappa-n_2^{-1})\{\tr({\boldsymbol\Omega}^2) + \tr^2(\boldsymbol\Omega)\} \right. \\
& \qquad \qquad + \left. (1-n_2^{-1})\{\tr({\mathbf{I}_K - {\boldsymbol\Omega}})^2 + \tr^2(\mathbf{I}_K - \boldsymbol\Omega)\} \right]^{-1}.
\end{align*}
Since $\widetilde{\boldsymbol{\Lambda}}$ is independently distributed to $\widetilde{\boldsymbol\zeta}_{1+} - \widetilde{\boldsymbol\zeta}_{2+}$, combining Eqn.~(\ref{eqn: supp_wishart_part}) into the representation of the test-statistic in~(\ref{eqn: supp_meat_rep}), and applying by Theorem 3.2.12 of \cite{muirhead2009aspects},
\begin{align*} 
\frac{ \nu n_2^{-2}  \mathcal{Z}^\top {\boldsymbol\Omega^\dagger}^{-1}  \mathcal{Z} } {\mathcal{Z}^\top \{(n_1+n_2-2)\mathcal{S}\}^{-1} \mathcal{Z}  }  \sim \chi^2_{\nu - K + 1}. 
\end{align*}
This implies 
\begin{align} 
 \frac{ \nu (n_1+n_2-2) \mathcal{Z}^\top {\boldsymbol\Omega^\dagger}^{-1}  \mathcal{Z} } {n_2^2(1/n_1 + 1/n_2)T_n  }  \sim \chi^2_{\nu - K + 1}. \label{eqn: supp_denom}
\end{align}
By the property of multivariate normal distribution, 
\begin{equation*} \label{eqn: supp_normal_part}
\mathcal{Z} = \sqrt{n_1}{\boldsymbol\Lambda^{\dagger}}^{-1/2} (\widetilde{\boldsymbol\zeta}_{1+} - \widetilde{\boldsymbol\zeta}_{2+}) \sim \text{N}_K(\sqrt{n_1}{\boldsymbol\Lambda^{\dagger}}^{-1/2}\boldsymbol\Delta, \mathbf{I}_K).
\end{equation*}
Suppose, that the full rank matrix ${\boldsymbol\Omega^\dagger}$ admits a spectral decomposition ${\boldsymbol\Omega^\dagger} = \sum_{k=1}^K d_k \boldsymbol{u}_k\boldsymbol{u}_k^\top$, with $\boldsymbol{u}_k^\top\boldsymbol{u}_j = \mathbb{I}(k = j)$  by Theorem 1 of \cite{baldessari1967distribution},
\begin{align}
 \mathcal{Z}^\top {\boldsymbol\Omega^\dagger}^{-1}  \mathcal{Z}  \sim {\displaystyle \sum_{k=1}^K d_k^{-1} \chi^2_1 \left (n_1 (\boldsymbol{u}_k^\top {\boldsymbol\Lambda^{\dagger}}^{-1/2} \boldsymbol\Delta)^2  \right )}. \label{eqn: supp_numer}
\end{align}
Dividing the left hand side of~(\ref{eqn: supp_numer}) with that of~(\ref{eqn: supp_denom}) we can see that our test statistic is approximately distributed as 
\begin{align*}
T_n &\overset{d}{=} \left\{\frac{\sum_{k=1}^K d_k^{-1} \chi^2_1 \left (n_1(\boldsymbol{u}_k^\top {\boldsymbol\Lambda^{\dagger}}^{-1/2} \boldsymbol\Delta)^2  \right )}{ \chi^2_{\nu - K + 1}/\nu}\right \} \\
& \times {(\kappa + 1 -2/n_2)}(1/\kappa + 1)^{-1}. 
\end{align*}
This completes the proof of Theorem.  

\subsection{Proof of Proposition~\ref{thm: power_egnunknown}}

Proof of Proposition~\ref{thm: power_egnunknown} directly follows from Theorem~\ref{thm: power_egnknown}, and by noticing that i) the degrees of freedom $\nu$ of the chi-square random variable in the denominator goes to $\infty$ as $n_1, n_2 \to \infty$,  and ii) $\chi^2_{\nu - K + 1}/\nu \to 1$ as $\nu \to \infty$ almost surely.

\section{Supplementary analysis for Azillect and SURE-PD3}\label{sec:appendixefunctions}

Table~\ref{tab:azillect_baseline} and \ref{tab:baseline_surepd3} presents the comparison of key PD related characteristics for the treatment and the placebo group at the baseline. The tables suggest that the two groups are homogeneous with respect to the baseline characteristics.

\begin{table}
\centering
\begin{tabular}{lllrrrr}
\hline \hline
 &  &  & \multicolumn{1}{r}{\begin{tabular}[r]{@{}l@{}}Placebo\\ $n = 126$\end{tabular}} &  &  & \multicolumn{1}{r}{\begin{tabular}[r]{@{}r@{}}Rasagiline\\ $n=118$\end{tabular}} \\ \hline
{ Age, in years, mean (SD)} &  &  & { $65.4 \,\,(8.81)$} &  &  & { $67.4 \,\,(8.96)$} \\
{ $\geq$ 65 years, $n (\%)$} &  &  & { $79 \,\,(63\%)$} &  &  & { $79 \,\,(67\%)$} \\
{ Female, $n (\%)$} &  &  & { $72 \,\,(57\%)$} &  &  & { $65 \,\,(55\%)$} \\
{ Duration of Parkinson's disease, years; mean (SD)} &  &  & { $1.47 \,\,(1.22)$} &  &  & { $1.81 \,\,(1.93)$} \\
{ Modified Hoehn \& Yahr stage, mean (SD)} &  &  & { $2.15 \,\,(0.61)$} &  &  & { $2.18 \,\,(0.63)$} \\
{ MDS-UPDRS Part II + Part III total score, mean (SD)} &  &  & { $33.8 \,\,(14.43)$} &  &  & { $34.4 \,\,(16.87)$} \\
{ MDS-UPDRS Part I total score, mean (SD)} &  &  & { $5.7 \,\,(3.58)$} &  &  & { $5.5 \,\,(3.82)$} \\
{ MDS-UPDRS Part II total score, mean (SD)} &  &  & { $7.0 \,\,(4.64)$} &  &  & { $7.2 \,\,(5.46)$} \\
{ MDS-UPDRS Part III total score, mean (SD)} &  &  & { $26.8 \,\,(11.59)$} &  &  & { $27.2 \,\,(13.74)$} \\
{ MDS-UPDRS tremor score, mean (SD)} &  &  & { $5.8 \,\, (4.64)$} &  &  & { $5.8 \,\,(4.51)$} \\
{ MDS-UPDRS bradykinesia score, mean (SD)} &  &  & { $11.6 \,\,(5.92)$} &  &  & { $12.3 \,\,(7.03)$} \\
{ MDS-UPDRS rigidity score, mean (SD)} &  &  & { $6.2 \,\,(3.33)$} &  &  & { $5.8 \,\,(3.33)$} \\
{ PDQ-39 summary index, mean (SD)} &  &  & { $11.94\,\, (11.49)$} &  &  & { $10.52 \,\,(10.00)$} \\
\hline \hline
\end{tabular}
\caption{Comparison of key PD related characteristics among the Rasagiline and Placebo group for patients in Azillect study. SD: standard deviation; MDS-UPDRS: Movement Disorder Society-Unified Parkinson's Disease Rating Scale; PDQ-39: Parkinson's Disease Questionnaire-39.}
\label{tab:azillect_baseline}
\end{table}

\begin{table}
    \centering
\begin{tabular}{lrr}
\hline \hline
% Characteristic & Inosine, $N=149^1$ & Placebo, $N=149^7$ \\ \hline  
 &  \multicolumn{1}{r}{\begin{tabular}[r]{@{}l@{}}Inosine\\ $n = 149$\end{tabular}} &  \multicolumn{1}{r}{\begin{tabular}[r]{@{}r@{}}Placebo\\ $n=149$\end{tabular}} \\ \hline 
Age, in years, mean (SD) & $63.0\,(9.71)$ & $63.6\,(9.39)$ \\
$\geq 65$ years, $n\,(\%)$ & $73\,(49 \%)$ & $74\,(50 \%)$ \\
 Female, $n$ (\%) & $80\,(54 \%)$ & $67\,(45 \%)$ \\
\multicolumn{3}{l}{ Race } \\
 \quad Asian & $2\,(1.4 \%)$ & $1\,(0.7 \%)$ \\
\quad Black or African American & $2\,(1.4 \%)$ & $0\,(0 \%)$ \\
 \quad Multiracial & $1\,(0.7 \%)$ & $1\,(0.7 \%)$ \\
 \quad White & $143\,(97 \%)$ & $145\,(99 \%)$ \\
 \multicolumn{3}{l}{ Ethnicity } \\
 \quad Hispanic or Latino & $5\,(3.4 \%)$ & $4\,(2.7 \%)$ \\
 \quad Not Hispanic or Latino & $144\,(97 \%)$ & $145\,(97 \%)$ \\
 Duration of Parkinson's disease, years; median (IQR) & $0.67\,(0.35$ -- $1.19)$ & $0.62\,(0.33$ -- $1.11)$ \\
 Modified S\&E ADL Scale score, mean (SD) & $94.1\,(4.82)$ & $94.1\,(4.83)$ \\
 Montreal Cognitive Assessment score, mean (SD) & $27.7\,(1.75)$ & $27.4\,(2.03)$ \\
 Neuro-QoL depression score, mean (SD) & $9.6\,(2.75)$ & $9.6\,(3.14)$ \\
 Modified Hoehn \& Yahr stage, mean (SD) & $1.67\,(0.51)$ & $1.75\,(0.44)$ \\
 MDS-UPDRS Part I+ II + Part III total score, mean (SD) & $31.4\,(12.49)$ & $34.1\,(12.64)$ \\
 MDS-UPDRS Part I total score, mean (SD) & $4.8\,(3.51)$ & $5.5\,(3.83)$ \\
 MDS-UPDRS Part II total score, mean (SD) & $5.3\,(3.91)$ & $6.1\,(4.42)$ \\
 MDS-UPDRS Part III total score, mean (SD) & $21.2\,(8.85)$ & $22.5\,(9.20)$ \\
 MDS-UPDRS tremor score, mean (SD) & $5.8\,(3.49)$ & $6.1\,(3.36)$ \\
 MDS-UPDRS bradykinesia score, mean (SD) & $9.9\,(5.50)$ & $10.7\,(5.57)$ \\
\hline
\end{tabular}
    \caption{Comparison of key PD related characteristics among the Inosine and Placebo group for patients in SURE-PD3 study. SD: standard deviation; MDS-UPDRS: Movement Disorder Society-Unified Parkinson's Disease Rating; Modified S\&E ADL: The modified Schwab and England Activities of Daily Living Scale; Neuro-QoL: The quality of life in neurological disorders assessment.}
    \label{tab:baseline_surepd3}
\end{table}

\begin{figure}
    \centering    \subfloat{\includegraphics[scale= 0.5]{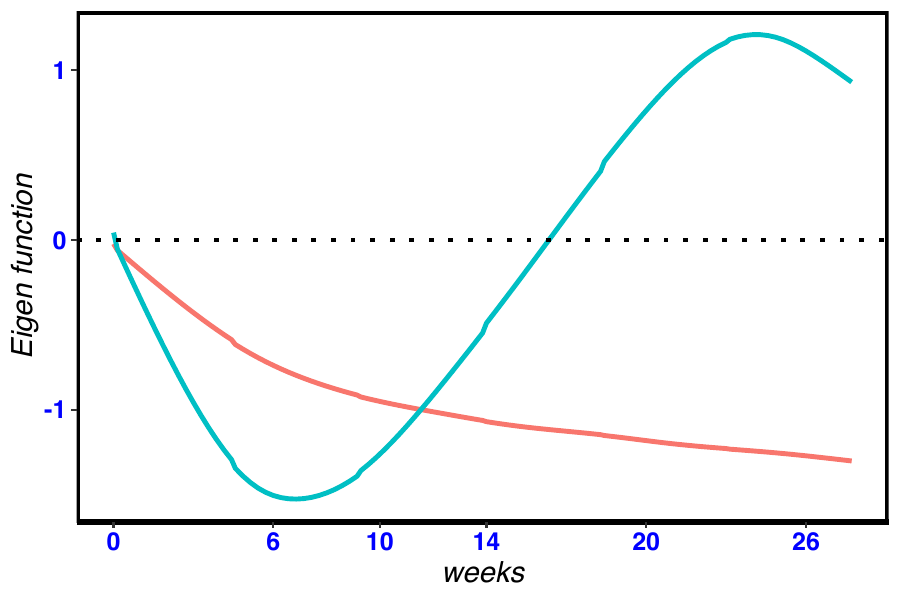}} \quad 
        \subfloat{\includegraphics[scale= 0.5]{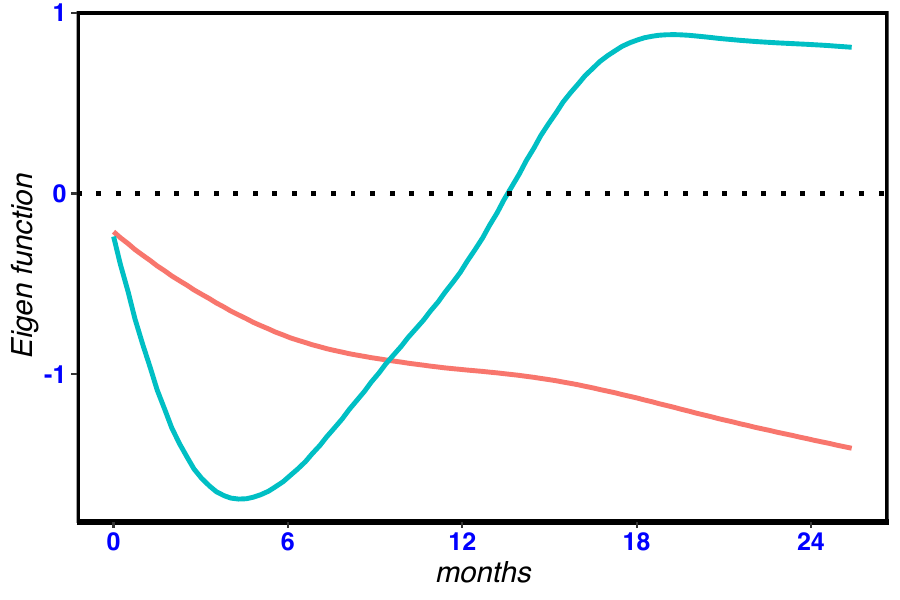}}
    \cprotect\caption{Estimated eigenfunction from the Azillect (left panel) and SURE-PD3 (right panel) study, based on a PVE of $95\%$.}
    \label{fig: efunctions}
\end{figure}

In Figure~\ref{fig: efunctions}, we present the first two eigenfunctions obtained by conducting fPCA in the Azillect and SURE-PD3 study, which are used to build the projection-based test and power analysis in Section~\ref{sec: realdata}. 

\end{document}